\documentclass[12pt,a4paper,epsf]{article}
\usepackage{epsfig}
\usepackage{amssymb}
\usepackage{graphicx}
\usepackage{tikz}
\usepackage{xcolor}
\usepackage{subfigure}
\usepackage[font=scriptsize]{caption}
\usepackage{latexsym}
\usepackage{framed}
\usepackage{amsmath}
 \usepackage{booktabs}
 \usepackage{multirow}
\usepackage{float}
\usepackage{cite}
\usepackage{hyperref}
\hypersetup{
    colorlinks,
    linkcolor={red!50!black},
    citecolor={blue!50!black},
    urlcolor={blue!80!black}
}

\begin{document}

\baselineskip 6mm
\renewcommand{\thefootnote}{\fnsymbol{footnote}}

\renewcommand{\thetable}{\Roman{table}}
\newcommand{\nc}{\newcommand}
\newcommand{\rnc}{\renewcommand}



\newcommand{\tcb}{\textcolor{blue}}
\newcommand{\tcr}{\textcolor{red}}
\newcommand{\tcg}{\textcolor{green}}


\def\beq{\begin{equation}}
\def\eeq{\end{equation}}
\def\ba{\begin{array}}
\def\ea{\end{array}}
\def\bea{\begin{eqnarray}}
\def\eea{\end{eqnarray}}
\def\nn{\nonumber}


\def\CMP{Commun. Math. Phys.~}
\def\JHEP{JHEP~}
\def\Pre{Preprint}
\def\PRL{Phys. Rev. Lett.~}
\def\PR {Phys. Rev.~}
\def\CQG {Class. Quant. Grav.~}
\def\PL {Phys. Lett.~}
\def\NP {Nucl. Phys.~}

\def\G{\Gamma}

\def\S{{\bf S}}
\def\C{{\bf C}}
\def\Z{{\bf Z}}
\def\R{{\bf R}}
\def\N{{\bf N}}
\def\M{{\bf M}}
\def\P{{\bf P}}
\def\bm{{\bf m}}
\def\bn{{\bf n}}

\def\CA{{\cal A}}
\def\CB{{\cal B}}
\def\CC{{\cal C}}
\def\CD{{\cal D}}
\def\CE{{\cal E}}
\def\CF{{\cal F}}
\def\CM{{\cal M}}
\def\CG{{\cal G}}
\def\CI{{\cal I}}
\def\CJ{{\cal J}}
\def\CL{{\cal L}}
\def\CK{{\cal K}}
\def\CN{{\cal N}}
\def\CO{{\cal O}}
\def\CP{{\cal P}}
\def\CQ{{\cal Q}}
\def\CR{{\cal R}}
\def\CS{{\cal S}}
\def\CT{{\cal T}}
\def\CV{{\cal V}}
\def\CW{{\cal W}}
\def\CX{{\cal X}}
\def\CY{{\cal Y}}
\def\We{{W_{\mbox{eff}}}}


\newcommand{\p}{\partial}
\newcommand{\bp}{\bar{\partial}}

\newcommand{\half}{\frac{1}{2}}

\newcommand{\bfalpha}{{\mbox{\boldmath $\alpha$}}}
\newcommand{\bfbeta}{{\mbox{\boldmath $\beta$}}}
\newcommand{\bfgamma}{{\mbox{\boldmath $\gamma$}}}
\newcommand{\bfmu}{{\mbox{\boldmath $\mu$}}}
\newcommand{\bfpi}{{\mbox{\boldmath $\pi$}}}
\newcommand{\bfvarpi}{{\mbox{\boldmath $\varpi$}}}
\newcommand{\bftau}{{\mbox{\boldmath $\tau$}}}
\newcommand{\bfeta}{{\mbox{\boldmath $\eta$}}}
\newcommand{\bfxi}{{\mbox{\boldmath $\xi$}}}
\newcommand{\bfkappa}{{\mbox{\boldmath $\kappa$}}}
\newcommand{\bfepsilon}{{\mbox{\boldmath $\epsilon$}}}
\newcommand{\bfTheta}{{\mbox{\boldmath $\Theta$}}}

\newcommand{\bz}{{\bar{z}}}

\newcommand{\dalpha}{\dot{\alpha}}
\newcommand{\dbeta}{\dot{\beta}}
\newcommand{\blambda}{\bar{\lambda}}
\newcommand{\btheta}{{\bar{\theta}}}
\newcommand{\bsigma}{{{\bar{\sigma}}}}
\newcommand{\bepsilon}{{\bar{\epsilon}}}
\newcommand{\bpsi}{{\bar{\psi}}}


\def\ct{\cite}
\def\la{\label}
\def\eq#1{(\ref{#1})}


\def\a{\alpha}
\def\b{\beta}
\def\g{\gamma}
\def\G{\Gamma}
\def\d{\delta}
\def\D{\Delta}
\def\ep{\epsilon}
\def\e{\eta}
\def\ph{\phi}
\def\Ph{\Phi}
\def\ps{\psi}
\def\Ps{\Psi}
\def\k{\kappa}
\def\l{\lambda}
\def\L{\Lambda}
\def\m{\mu}
\def\n{\nu}
\def\th{\theta}
\def\Th{\Theta}
\def\r{\rho}
\def\s{\sigma}
\def\S{\Sigma}
\def\ta{\tau}
\def\o{\omega}
\def\O{\Omega}
\def\pr{\prime}


\def\half{\frac{1}{2}}

\def\goto{\rightarrow}

\def\na{\nabla}
\def\grad{\nabla}
\def\curl{\nabla\times}
\def\div{\nabla\cdot}
\def\pa{\partial}

\def\bra{\left\langle}
\def\ket{\right\rangle}
\def\lb{\left[}
\def\lc{\left\{}
\def\ls{\left(}
\def\lp{\left.}
\def\rp{\right.}
\def\rb{\right]}
\def\rc{\right\}}
\def\rs{\right)}
\def\cl{\mathcal{l}}

\def\vac#1{\mid #1 \rangle}

\def\td#1{\tilde{#1}}
\def\check{ \maltese {\bf Check!}}


\def\Tr{{\rm Tr}\,}
\def\det{{\rm det}\,}


\def\bc#1{\nnindent {\bf $\bullet$ #1} \\ }
\def\ch {$<Check!>$ }
\def\ss {\vspace{1.5cm}}

\begin{titlepage}

\hfill\parbox{5cm} { }

\hskip1cm


\vspace{10mm}

\begin{center}
{\Large \bf ``Striped"  Rectangular Rigid Box with Hermitian and non-Hermitian $\mathcal{PT}$  Symmetric Potentials  }

\vskip 0.8 cm
  {
 
  Shailesh Kulkarni\footnote{e-mail : shailesh@physics.unipune.ac.in}
  {\it and } 
Rajeev K. Pathak\footnote{e-mail : snehalandrajeev@gmail.com }  
  }

\vskip 0.2cm

{\it Department of Physics, Savitribai Phule Pune University, Ganeshkhind, Pune, 411007, India }\\

\end{center}

\thispagestyle{empty}

\vskip0.8cm


\centerline{\bf ABSTRACT} \vskip 2mm
\noindent Eigenspectra of a spinless quantum particle trapped inside a rigid, rectangular, two-dimensional (2D) box subject to diverse inner potential distributions are investigated under hermitian, as well as non-hermitian  antiunitary $\mathcal{PT}$  (composite parity and time-reversal)  symmetric regimes.   Four sectors or “stripes” inscribed in the rigid box comprising contiguously conjoined parallel rectangular segments with one side equaling the entire width of the box are studied.  The stripes encompass piecewise constant potentials whose exact, complete energy eigenspectrum is obtained employing matrix mechanics.  Various striped potential compositions, viz.  real valued ones  in the hermitian regime as well as complex, non-hermitian but $\mathcal{PT}$ symmetric ones are considered separately and in conjunction, unraveling among typical lowest lying eigenvalues, retention and breakdown scenarios engendered by the $\mathcal{PT}$ symmetry, bearing upon the strength of non-hermitian sectors.  Some states exhibit a remarkable crossover of symmetry ‘making’ and ‘breaking’: while a broken $\mathcal{PT}$ gets reinstated for an energy level, {\it higher} levels may couple to continue with symmetry breaking.  Further, for a charged quantum particle a $\mathcal{PT}$ symmetric electric field, furnished with a striped potential backdrop, also reveals peculiar retention and breakdown $\mathcal{PT}$  scenarios.  Depictions of prominent probability redistributions relating to various potential distributions both under norm conserving unitary regime for hermitian Hamiltonians and non-conserving ones post $\mathcal{PT}$  breakdown are presented.
\vspace{1cm}
\noindent 
\vspace{2cm}


\end{titlepage}

\renewcommand{\thefootnote}{\arabic{footnote}}
\setcounter{footnote}{0}


\section{Introduction}

\indent
 Among  the  few  problems  that  pragmatically  merit  exact  quantum  mechanical  solutions,  the ‘rigid-box’   problems   form   a   special   class   signifying,   contingent   on   dimensionality   and symmetry, a quantum particle confined within some standard geometries such as line segments, rectangular  and  circular  areas,  rectangular  parallelepipeds  (cuboids),  cubes,  spheroids,  spheres and  thus  forth.    In  all  these  cases,  as  is  well-known,  appropriately  chosen  coordinate  systems permit   separation   of   variables   in   the   configuration   space,   yielding   analytical   solutions. Following its inception by McDonald and Kaufman \cite{MCDKAUF79}, ‘Quantum Billiards’ (QB), that has emerged a generic term for analyzing the dynamics of a particle in confined spaces, has stimulated  widespread  interest  over  the  past  four  decades \cite{RMP17, ROB1, ROB2, ROB3,ROB3B, ROB4, ROB5}. Although  even  a  slightest  deviation  from  standard  shapes  could render  a  systematically  tractable  solution  impossible,  cogent 
 numerical  methods  to  tackle  such situations   have   been   developed \cite {KOSZ97, KAUF99}.  Interestingly,   particles   moving   in polygonal enclosures whose interior angles are rational multiples of $\pi$ were observed to exhibit `pseudo-integrability'  embodying   tenets   of   quantum   chaos \cite{RICH81}.  Moreover,   some customized  mesoscopic  semiconductor  quantum  dots  were  experimentally  observed  to  exhibit salient QB characteristics in terms of forward and back-scattered electron wave-packets and their revivals \cite{BERRY1994}.        Later,  a  case  strikingly  akin  to  this  effect  was  impeccably  solved {\it exactly} by Robinett \cite{ROB3}, for a 2D rigid circular box augmented by an infinite thin barrier or baffle  introduced  along  one  of  its  radii  constituting  quantum  billiards.    Noticeably,  QB  were demonstrated  to  furnish  a  manifest  natural  connection  between  the  complementary  phenomena of classical and quantum chaos \cite{RMP17, ROB1}.  

Meanwhile,  Bender  and  Boettcher  \cite{BEND98},  through  their  pioneering  article  in  1998,  launched  the  exotic theme of  $\mathcal{PT}$ Symmetry, signifying invariance  of the Hamiltonian under the composite, discrete symmetry operations of Parity ($\mathcal{P}$) and  Time-reversal ($\mathcal{T}$ ), irrespective of  their  order  of  application,  for  a  wide  variety  of  systems.    Introduction  of  $\mathcal{PT}$ symmetry,  a remarkable non-Hermitian extension of Quantum Mechanics, has evoked inexorable intrigue and has  stimulated  a  multitude  of  theoretical  \cite{BEND98,    BEND19}  as  well  as  experimental \cite{CHRIS2018, RELG17}  ventures.    When  in  a  Hamiltonian  system,  a  perfect  balance  is  struck between gain and loss mechanisms, $\mathcal{PT}$ symmetry manifests through the corresponding potential (energy)  terms. $\mathcal{PT}$ symmetry  also  got  introduced  in  confined  quantum  systems  by  several workers:  notably, Bittner {\it et al}.  \cite{BITT2012} examined a two-state Hamiltonian of a dissipative microwave  billiard  in  the  neighborhood  of  an  exceptional  point,  bringing  forth  a  peculiar $\mathcal{PT}$ symmetry.   Further,   Dasarathy {\it et   al}.   \cite{DASA2013}  imposed $\mathcal{PT}$ symmetry   not   by parameterizing the Hamiltonian, but rather implicitly, through the {\it boundary conditions} over the standard 1D rigid box segment and established that the ensuing dynamics invariably  conformed to  a  robust $\mathcal{PT}$   symmetry.  In  their  recent  experiments,  Gu  and  coworkers  \cite{ZGUN16} ingeniously  demonstrated  incorporation  of   $\mathcal{PT}$   symmetry  in  a  ``Stripe-LASER"  waveguide  with its pumped part functioning  as `gain', and the other unpumped one serving as `loss', due to  the  intrinsic  absorption  of  a  deliberately  introduced  dye.    Interestingly,  Kreibich  \cite{KREI15} presented  a  concrete  scenario  for  realizing $\mathcal{PT}$ symmetry  for  particle  currents  in  Bose-Einstein condensates,  simulating  a  two-mode  model  system  embedded  in  a  larger {\it hermitian}  system having  at  least  four  additional  coupled  ‘reservoir’  wells  equipped  with  a  balanced  gain-loss mechanism.      For   discrete   confined   systems,   Musslimani {\it et  al}.   \cite{MUSS08}   conclusively demonstrated  that $\mathcal{PT}$ symmetry  can  support  soliton  solutions  in  1D  as  well  as  2D  nonlinear optical lattices. $\mathcal{PT}$ symmetry thus is not limited to a   theoretical abstraction, but also presents tangible perspectives amenable to direct experimentation \cite{BEND19, RELG17, CHRIS2018}.  

Induction  of $\mathcal{PT}$ symmetry  in  the  ‘particle-in-a-box’  context  was  likewise  carried  out,  in particular, by Yusupov et al.  \cite{YUSU18}, who studied the
quantum dynamics in a 1D box, driven by  a $\mathcal{PT}$ symmetric  complex  potential,  with  impulsive,  non-Hermitian,  delta-function  ‘kicks’, and   observed   that   the   otherwise   sustained $\mathcal{PT}$ symmetry   broke   down   beyond   certain characteristic  threshold  strengths  of  the kicking  parameter.   Interestingly,  notwithstanding  the standard  constraints  for  the  elementary  1D  rigid  box  problem,  Adamu  \cite{ADAM14},  by  a  direct {\it post facto}  imposition of the $\mathcal{PT}$ symmetric {\it boundary conditions} on the eigenfunctions, obtained  a  distinct  class  of  eigenfunctions  associated  with the  same  eigenvalues  as  that  of  the  usual  1D box.    
Around  the  same  time,  Fernàndez  and  Garcia \cite{FERN14}  applied  perturbation  theory  to  a 2D  rigid  square  box  incorporating  some  model  parameterized $\mathcal{PT}$ symmetric  potentials,  that exhibited retention of $\mathcal{PT}$ symmetry below some specific threshold parameter-values. 

For  yet  another  confined  2D  system  albeit  with  circular  geometry,  Agarwal  and  others \cite{AGAR15}  presented  exact  solutions  to  typified $\mathcal{PT}$ symmetric  potentials  with  sinusoidally oscillating  azimuth,  revealing  some  counterintuitive $\mathcal{PT}$ breakdown  scenarios,  e.g.  raising  the hermitian component sometimes resulted in accelerating, rather than deterring, breaching of $\mathcal{PT}$.  Further,  in  another  exploration  on $\mathcal{PT}$ applied  to  2D, finite {\it discrete}  latticework, these workers \cite{AGAR18} demonstrated that for a set of coupled chains in 2D with only two balanced gain-loss $\mathcal{PT}$ sites  interspersed, the  transition  threshold  could  be  markedly  enhanced  as  a function of the coupling, providing a handle on  tune-ability.  Incidentally, for a 1D  rigid box, a family  of $\mathcal{PT}$ symmetric  complex  potentials  isospectral  with  its eigenspectrum  was  obtained \cite{CHER13}.  

The  foregoing  discussion  emphasizes  that  quantum  particles  in  entrenched,  or  confined, quantum  billiards type  configurations  could  exhibit  uncommon  dynamics,  accentuated  with application  of $\mathcal{PT}$  symmetry.    Subscribing  to  this  premise,  the  present  article  is  aimed  at systematically  studying  the  behavior  of  a  spinless particle  in  a  `striped'  two dimensional rectangular  rigid  quantum  box,  a  special  type  of  quantum  billiards  whose  {\it exact}  quantum mechanical energy eigenspectrum will be obtained.  Piecewise constant potentials are introduced along  the  breadth  of  the  box,  in  chosen  sectors  or stripes,  where  some  or  all  of  them  are selectively rendered hermitian and/or $\mathcal{PT}$ symmetric, the latter accomplished by introduction of balanced gain  and loss sectors.  We delineate onset of $\mathcal{PT}$ symmetry breaking at characteristic  exceptional points and bring  forth  the consequent  mutations in the probability density.   While the present venture evidently provides {\it exact} solutions to a model quantum mechanical situation, a  pertinent  physical  problem  that  could  serve  as  its  excellent  prototype  would  be  a  three dimensional   tubular   wave-guide   with   a   uniform   rectangular   normal   cross-section,   with propagation of the particle or light flux in the longitudinal positive $z$ direction, and a suitably tune-able lateral  beam  distribution  in  the $x-y$ plane:  for  a  1D  analogue,    reference  may  be  made  to  the
work of Moiseyev and others \cite{MOIS2008}, who studied a waveguide with a piecewise constant {\it complex}  refractive  indices.      In  the  present  problem,  preemptively  factoring  out  the  plane-wave propagation along positive $z$, we focus on the wave function projected onto the plane, within the ambit of  the  rectangle.    It  will  turn  out  that  the  exactly  solvable  2D  rectangular  rigid  box,  in conjunction with a stipulated $\mathcal{PT}$ symmetric potential distribution in the interior also selectively exhibits  sustenance  and  abrogation  of  the  symmetry consequent  to  the  interplay  between  the relative strengths of the intervening potentials.

\section{Two dimensional rectangular rigid box with striped potential distribution: exact solutions}  
\begin{figure}[h]
\vspace*{-20mm}
\centering
\begin{tikzpicture}[x=0.45pt,y=0.4pt,yscale=-1,xscale=1]

\draw [line width=1.5]    (183,531) -- (596,530.11) ;
\draw [shift={(599,530.1)}, rotate = 539.88] [color={rgb, 255:red, 0; green, 0; blue, 0 }  ][line width=1.5]    (14.21,-4.28) .. controls (9.04,-1.82) and (4.3,-0.39) .. (0,0) .. controls (4.3,0.39) and (9.04,1.82) .. (14.21,4.28)   ;
\draw [line width=1.5]    (186,531.1) -- (183.02,169.1) ;
\draw [shift={(183,166.1)}, rotate = 449.53] [color={rgb, 255:red, 0; green, 0; blue, 0 }  ][line width=1.5]    (14.21,-4.28) .. controls (9.04,-1.82) and (4.3,-0.39) .. (0,0) .. controls (4.3,0.39) and (9.04,1.82) .. (14.21,4.28)   ;
\draw [color={rgb, 255:red, 0; green, 0; blue, 0 }  ,draw opacity=1 ][line width=1.5]    (183,232.1) -- (439,231.1) ;
\draw [line width=1.5]    (440,230.1) -- (440,530.1) ;
\draw [line width=1.5]    (184,324.1) -- (440,323.1) ;
\draw [line width=1.5]    (185,435.1) -- (440,433.1) ;
\draw    (185,379.1) -- (441,379.1) ;

\draw (721,21) node    {$$};
\draw (435,538.5) node [anchor=north west][inner sep=0.75pt]    {$a$};
\draw (152,226.5) node [anchor=north west][inner sep=0.75pt]    {$b$};
\draw (150,317.5) node [anchor=north west][inner sep=0.75pt]  [color={rgb, 255:red, 0; green, 0; blue, 0 }  ,opacity=1 ]  {$b_{3}$};
\draw (148,428.5) node [anchor=north west][inner sep=0.75pt]    {$b_{1}$};
\draw (83,372.5) node [anchor=north west][inner sep=0.75pt]    {$b_{2} =b/2$};
\draw (169,532.5) node [anchor=north west][inner sep=0.75pt]    {$0$};
\draw (148,164.5) node [anchor=north west][inner sep=0.75pt]    {$y$};
\draw (583,536.5) node [anchor=north west][inner sep=0.75pt]    {$x$};
\draw (217,275.1) node [anchor=north west][inner sep=0.75pt]  [font=\small] [align=left] {{\small region-IV}};
\draw (217,347.1) node [anchor=north west][inner sep=0.75pt]   [align=left] {{\small region-III}};
\draw (219,403.1) node [anchor=north west][inner sep=0.75pt]   [align=left] {{\small region-II}};
\draw (222,475.1) node [anchor=north west][inner sep=0.75pt]   [align=left] {{\small region-I}};
\draw (349,278.5) node [anchor=north west][inner sep=0.75pt]  [font=\small]  {$V_{4}$};
\draw (351,352.5) node [anchor=north west][inner sep=0.75pt]  [font=\small]  {$V_{3}$};
\draw (350,404.5) node [anchor=north west][inner sep=0.75pt]  [font=\small]  {$V_{2}$};
\draw (354,478.5) node [anchor=north west][inner sep=0.75pt]  [font=\small]  {$V_{1}$};

\end{tikzpicture}
\caption{A schematic “striped” rectangular box in the $x-y$ plane, with $0< x < a,  0 <y < b$; partitioned with four rectangular sectors I, II, III and IV respectively bearing the piecewise constant, finite potentials $V_1, V_2, V_3$ and $V_4$. The partitions are symmetric with respect to the median $y = b/2 \equiv b_2$, whence, $b_3-b_2 = b_2-b_1 < b/2$.  Outside the box, throughout, the potential is $+\infty$.}
\label{fig:Figure-1}
\end{figure}
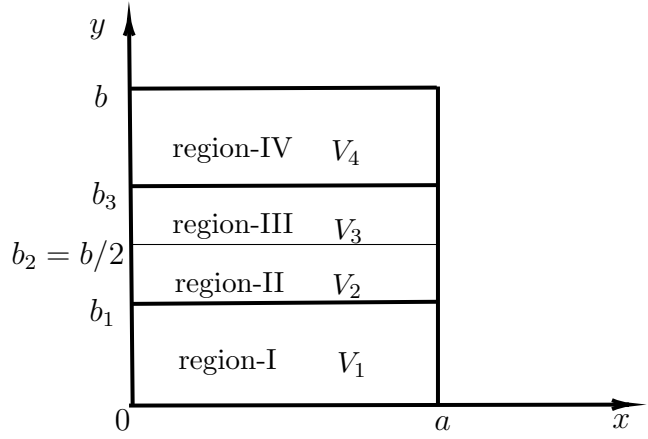

 We consider a rigid rectangular box in the Cartesian $x-y$ plane, bounded by $\displaystyle 0 \le x \le a ,  0 \le y \le b$, partitioned with piecewise constant potentials 
$V_1, V_2, V_3$ and $V_4$ in contiguous rectangular sectors or stripes throughout their breadth, equaling the breadth of the rigid box, parallel to the $x$-axis.
 The inside potentials are finite with possible finite discontinuities at each interface within the box. The rectangular  sectors,  i.e.  the stripes are  symmetric around the $y$-median as depicted in the accompanying schematic Figure-\ref{fig:Figure-1}.   The region outside the box, of course, is impenetrable, with positive infinite potential. 

The potential $V(x,y)$ within the rigid box is designated as follows: 
$$
V(x,y)=
\begin{cases}
V_4,  \ b_3 < y <b \\
V_3, \ b_2<y<b_3 \\
V_2, \ b_1<y<b_2 \\
V_1, \ 0<y<b_1
\end{cases}
$$
with  $0< x < a$  throughout and $V(x,y) \rightarrow +\infty$, outside the rectangle.   
Note that within the box, the only explicit dependence of the potential $V(x,y)$ is on the $y$-coordinate; therefore inside, $V(x,y) \equiv V(y)$
 alone, a fact that will be exploited shortly.  The stationary states for a scalar (spinless) particle of mass $\mu$ 
trapped inside the planar $2$D rectangular geometry are the solutions of the time-independent Schrödinger equation 
\begin{equation}
\hat{H}\psi(x,y) \equiv -\frac{\hbar^2}{2\mu}\Big[\nabla^{2}_{2D} + V(x,y)\Big]\psi(x,y) = E \psi(x,y) 
\label{schro1}
\end{equation}
where $\nabla^2_{2D}$ is $2$D Laplacian operator and $\psi(x,y)$ designating the time-independent energy eigenfunctions associated with energy eigenvalues $E$ of the Hamiltonian operator $\hat{H}$. Let us recall that for the standard, rudimentary text book case of a particle entrapped in the rigid box, but otherwise free inside, 
i.e. $V(x,y) = 0 $ throughout the box, the corresponding eigenfunctions $\psi^{(0)}_{n_x,n_y}(x,y)$  factorize themselves as $\psi^{(0)}_{n_x,n_y}(x,y) = u_{n_x}(x) u_{n_y}(y)$ within the box and identically vanish outside the rectangle. The quantum numbers $n_x$ and $n_y$ take values $1,2,\cdots$. The normalized forms of the factor functions are 
$$
u_{n_x}(x) = 
\begin{cases}
\sqrt{\frac{2}{a}}\sin\left(\frac{n_x \pi x}{a}\right),  0\le x \le a \\
0, \text{otherwise}
\end{cases}
$$
and 
$$
u_{n_y}(y) =
\begin{cases}
\sqrt{\frac{2}{b}}\sin\left(\frac{n_y \pi y}{b}\right),  0 \le  y \le b \\
0, \text{otherwise}
\end{cases}
$$
while the energy eigenvalues emerge as 
$$E^{(0)}_{n_x,n_y} = \frac{\pi^2 \hbar^2}{2\mu}\Big(\frac{n_{x}^{2}}{a^2}+\frac{n_{y}^2}{b^2}\Big)$$
Note that degeneracies, including accidental degeneracies, could occur in this bound state two dimensional (2D) 
problem. To solve for the eigenspectrum of the striped box, we expand the wave function $\Psi(x,y)$ {\it vide} the completeness of the product functions
$u_{n_x}(x) u_{n_y}(y)$ :
\begin{equation}
\psi(x,y) = \sum_{n_x, n_y=1}^{\infty} C_{n_{x},n_{y}}u_{n_x}(x) u_{n_y}(y) 
\end{equation}
Substituting the above in Eq.~(\ref{schro1}),  one is led to 
\begin{eqnarray}
\sum_{n_x=1}^{\infty}u_{n_x}(x)\sum_{n_y=1}^{\infty}
\Big[\frac{\pi^2\hbar^2}{2\mu}\Big(\frac{n_{x}^{2}}{a^2}+\frac{n_{y}^2}{b^2}\Big)+V(y)\Big]C_{n_{x},n_{y}}u_{n_y}(y) = 
E\sum_{n_x=1}^{\infty}u_{n_x}(x)\sum_{n_y=1}^{\infty}C_{n_{x},n_{y}}u_{n_y}(y);
\end{eqnarray}
owing to the explicit dependence of the potential exclusively on the $y$-coordinate, within the rectangle.  
This leads to the separation of the $x$-solution altogether, whence, exploiting the linear independence of the functions 
$\{u_{n_x}(x)\}$ one arrives at
\begin{equation}
\sum_{n_y=1}^{\infty}
\Big[\frac{\pi^2\hbar^2}{2\mu}\Big(\frac{n_{x}^{2}}{a^2}+\frac{n_{y}^2}{b^2}\Big)+V(y)-E\Big]C_{n_{x},n_{y}}u_{n_y}(y) = 0. \label{schro2}
\end{equation}
Since the $x$-dependence separates out (analogous to that coordinate being “cyclic” in the Hamiltonian defined over the box-region), 
we can designate the $n_x$ quantum number a definite, fixed value, say $n_x \equiv n_{x}^{(0)}$ for the $x$-solution, which is then denoted by 
$u_{n_{x}^{(0)}}(x)$.
Next, we substitute the appropriate piecewise constant forms for $V(y)$ in the four sectors, clamp the $n_x$ quantum number 
at the chosen $n_{x}^{(0)}$ value,  pre-multiply the above Eq.~(\ref{schro2}) throughout by $u^{*}_{n'_{y}}(y)$ and integrate over the $y$ in the domain $[0,b]$. 
Implementing this in sequential conjunction and harnessing the orthonormality of $u_{n_y}(x)$ functions the procedure results in the following connection: 
\begin{eqnarray}
\sum_{n_{y}=1}^{\infty}\Big\{\frac{\pi^2\hbar^2}{2\mu}\Big[\frac{(n_{x}^{(0)})^{2}}{a^2}+\frac{n_{y}^2}{b^2}\Big]-E\Big\}C_{n_{y}}^{n^{(0)}_{x}} \delta_{n'_{y}n_{y}}
+\frac{2}{b}\sum_{n_{y}=1}^{\infty}\Big( \sum_{i=1}^{4} \ V_i  \ \int_{b_{i-1}}^{b_i} \ dy \ \phi(y)\Big)C_{n_{y}}^{n^{(0)}_{x}} =0 
\end{eqnarray}
where $b_0 \equiv 0$ and $b_4 \equiv b$ and $\phi(y)= \sin \Big(\frac{n'_{y}\pi y}{b}\Big)\sin\Big(\frac{n_{y}\pi y}{b}\Big)$. The coefficients have been relabeled as 
$C_{n^{(0)}_{x},n_{y} } \equiv C_{n_{y}}^{n^{(0)}_{x}}$, reminding that only the discrete index $n_y$ actually varies, the index $n^{(0)}_{x}$ having been clamped.  The wave function then simplifies to
\begin{equation}
\psi(x,y) = u_{n^{(0)}_{x}}(x) \sum_{n_y =1}^{\infty} C_{n_{y}}^{n^{(0)}_{x}} u_{n_y}(y) \label{psi}
\end{equation}
 After evaluation of the integrals the above line of arguments engenders a matrix eigenvalue equation 
\beq
[M][C] = E [C]. \label{mateq}
\eeq
Here, the matrix $M = [M]_{n'_{y},n_y}$ is an infinite dimensional `square' matrix, labeled by the respective row and column indices $n'_{y}, n_y = 1,2,3\cdots$; while 
$C_{n_{y}}^{n^{(0)}_{x}} $ incarnates as a column vector. The matrix elements of $M$ bear the explicit form
\bea
M_{n_y,n_y} &=& \Big[\frac{\pi^2\hbar^2}{2\mu}\Big(\frac{(n_{x}^{(0)})^{2}}{a^2}+\frac{n_{y}^2}{b^2}\Big)\Big]
+\frac{1}{b}\sum_{i=1}^{3} b_i (V_i - V_{i+1}) + V_4  \nonumber\\ 
&&
+ \frac{1}{2\pi n_y}\Big[\sum_{i=1}^{3} (V_{i+1}-V_i)\sin\Big(\frac{2\pi n_{y}b_i}{b}\Big)\Big] ; \label{mateq1}
\eea
for the diagonal entries, and 
\bea
M_{n'_y,n_y} &=& \frac{1}{\pi(n'_{y}-n_y)}\Big[\sum_{i=1}^{3}(V_i - V_{i+1})\sin\Big(\frac{\pi (n'_{y}-n_y)b_i}{b}\Big) \Big] \nonumber \\  
&&
+\frac{1}{\pi(n'_{y}+n_y)}\Big[\sum_{i=1}^{3} (V_{i+1}-V_i)\sin\Big(\frac{\pi (n'_{y}+n_y)b_i}{b}\Big)\Big];  \label{mateq2}
\eea
for the off-diagonal entries ($n'_{y}\ne n_y)$. 
We shall put the foregoing analysis to test with a variegated set of hermitian, and non-hermitian but $\mathcal{PT}$
symmetric combinations, as carried out in the next section.
\section{Results and discussion}
To tackle the stationary-state problem for seeking the eigenvalues and eigenfunctions, we must adopt an appropriate scale.  Setting the Bohr radius 
$\displaystyle a_{B} \equiv \frac{4\pi \epsilon_{0}\hbar^2}{m_e e^2}$ (in usual notation) to represent a unit distance, and $\hbar$ and the electron mass, $m_e$ both numerically unity, the energy scale is set by the quantity $\frac{\hbar^2}{2m_e a_{B}^{2}}=13.605693  \ eV \equiv 1$  Rydberg energy.
Thus, we shall measure distances in units of Bohr radii and energies in units of Rydbergs.  We also have, for convenience (although not direly necessary), chosen the mass of the quantum particle $\mu$ to be the electron mass, $m_e$. The task now is to solve the matrix equation, Eq. (\ref{mateq}), 
with the prescriptions in Eqs. (\ref{mateq1}) and (\ref{mateq2}) above.  For nontrivial solutions, the secular equation, viz.  
$det [M - EI]=0$ (with $I$ denoting the identity matrix) must be satisfied, which determines the energy eigenvalues  $E$  and thereafter, corresponding to each eigenvalue, the column vector  $C$ formed of the expansion coefficients, readily synthesizes the wave function as stipulated by Eq. (\ref{psi}).  

Although in principle, the range spanned by the quantum numbers $n'_{y}, n_y$ goes from $1$ to $\infty$, in practice, a proper upper limit $n_{y}^{max}$ must be chosen for numerical implementation of the exact matrix-eigenvalue-equation, Eq.(\ref{mateq}).  It turns out that for an accurate evaluation the lowest $15$-odd energy eigenvalues for maximal magnitude of the $V_i$ (cf. Figure-\ref{fig:Figure-1}) $150 \ Ry$ energy units, the upper limit of $n_{y}^{max}= 30$ suffices for convergence to occur.   We have chosen a better stringent and robust upper limit of $n_{y}^{max}= 50$, for reaffirming convergence of the lowest eigenvalues, rendering the computations numerically completely unequivocal.  The convergence is markedly rapid, which may be attributed to the following: for sufficiently large values of  $n'_{y}$ and $n_y$, the diagonal elements of the Matrix $M$, (cf. Eq.(\ref{mateq1})) form the major contribution, as they scale as $\sim (n'_{y})^2$, 
while the off-diagonal ones far removed from the principal diagonal,  scale magnitude wise as $\sim 1/n'_{y}$ or $\sim 1/ n_y$ (cf. Eq.(\ref{mateq2})),  
endowing a relative significance of $\sim (n'_{y})^{3}$  to the diagonal elements over the far off-diagonals, thereby imparting a desirable characteristic for convergence.  

 If the origin is chosen to be the center of inversion, the parity operator $\mathcal{P}$, which is a linear and hermitian transformation maps a generic wave function  $\psi(\vec{r},t)$ in accord with  $\mathcal{P} \psi(\vec{r},t) =\psi(-\vec{r},t)$ whereas the antilinenear time-reversal $\mathcal{T}$ transformation accomplishes the mapping $\mathcal{T} \psi(\vec{r},t)=\psi^{*}(\vec{r},-t)$.  Further, since $\mathcal{P} = \mathcal{P}^{-1}$, for the present {\it spinless} case, $\mathcal{T} =\mathcal{T}^{-1}$ (while  incidentally, for half-odd-integral spins, due to Kramers's degeneracy, $\mathcal{T}^2 = -I$, the negative identity), in the coordinate representation for the position and canonical momentum operators respectively follow the operator identities: 
$$\mathcal{P}  \ \vec{r} \ \mathcal{P}^{-1}=-\vec{r}, \ \mathcal{T}  \ \vec{r} \ \mathcal{T}^{-1}=\vec{r}, \ \mathcal{P}  \ \hat{\vec{p}} \ \mathcal{P}^{-1}=-\hat{\vec{p}}, \ \mathcal{T}  \ \hat{\vec{p}}  \ \mathcal{T}^{-1} = -\hat{\vec{p}}$$
Further $[\mathcal{P},\mathcal{T}]=0$, and $\hat{H}$ is said to be $\mathcal{PT}$ symmetric if $[\hat{H},\mathcal{PT}]=0$.  This requires that for the single spinless particle \cite{KREI15} : 
\bea
(\mathcal{PT}) \hat{H} = \Big[\frac{(\hat{\vec{p}})^2}{2\mu} + V^{*}(-\vec{r})\Big] \mathcal{PT}  = \hat{H}(\mathcal{PT}), \label{PTeq}
\eea
since the antilinear $T$ symmetry deems the transformation for a general complex-valued, explicitly time independent potential $V$, as
$V\equiv Re(V) + i  \ Im(V) \rightarrow Re(V) -i \ Im(V) = V^{*}$. Further, for an eigenstate $|\psi\rangle$,

\beq
\hat{H}|\psi\rangle = E|\psi\rangle \implies \hat{H} \ \mathcal{PT} \ |\psi\rangle = E^{*} \ \mathcal{PT} \ |\psi\rangle; \label{PTeq2}
\eeq 
 thus retention of $\mathcal{PT}$ symmetry manifests in reality of the eigenvalues, while post-$\mathcal{PT}$ breakdown, energy eigenvalues must occur in complex conjugate-pairs. 
$\mathcal{PT}$ symmetry is still intriguing, since, especially for a multivariate problem, it is indiscernible {\it a priori}, exactly in which parameter-regime the symmetry will be broken at the ``exceptional points" and further, the  symmetry breaking is spontaneous \cite{BEND99}.  Conservation of currents from a general field-theoretic perspective was established for the $\mathcal{PT}$ operation by Alexandre et al. \cite{ALEX2017}. 
 
If it is required that the regions I through IV have $\mathcal{PT}$ symmetry incorporated in the potential (energy) part, 
with the $\mathcal{P}$(parity) operation carried out around the median  $y = b/2 \equiv b_2$, that is accomplished through $y-b/2 \rightarrow b/2-y$.  As is well-known, if the potential is completely real,  $\hat{H}$  is manifestly hermitian with no extra restrictions for reality of its associated eigenvalues.  However, when a real potential in some sector is accompanied by a complex valued or a pure imaginary one in another sector, real eigenvalues in the sustained, unbroken $\mathcal{PT}$  regime could result only when the  $Re(V)$ is an even, i.e. a symmetric function under spatial inversion (parity), while the $Im(V)$ is odd, i.e. antisymmetric. We shall impose this requirement on the general, complex-valued potential distributions chosen herein. 

For the rectangular $2$D rigid box studied herein we set $a=\sqrt{3}, b=\sqrt{2}$ both in the units of Bohr radii.  Such a choice of irrational distance units should keep the accidental degeneracies to the minimum, if not completely suppress them.  Further, $b_1$ and $b_3$, located symmetrically with respect to the median $b_2= b/2$ are set to: $b_1=0.4 b$ and $ b_3=0.6 b$. 
\subsection{Real valued striped potentials}
\begin{figure}[h]
  \includegraphics[width=0.3\columnwidth]{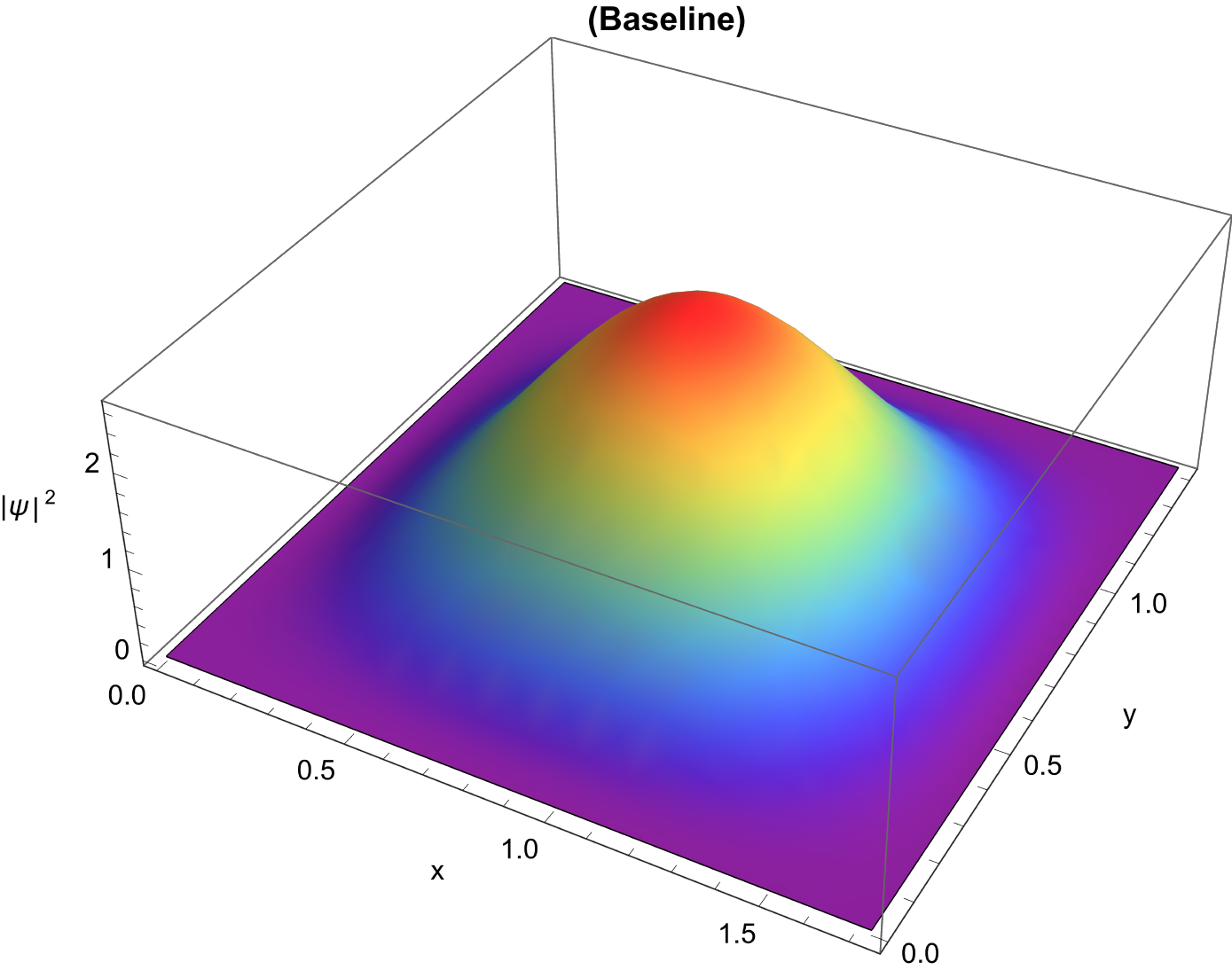}
 \includegraphics[width=0.3\columnwidth]{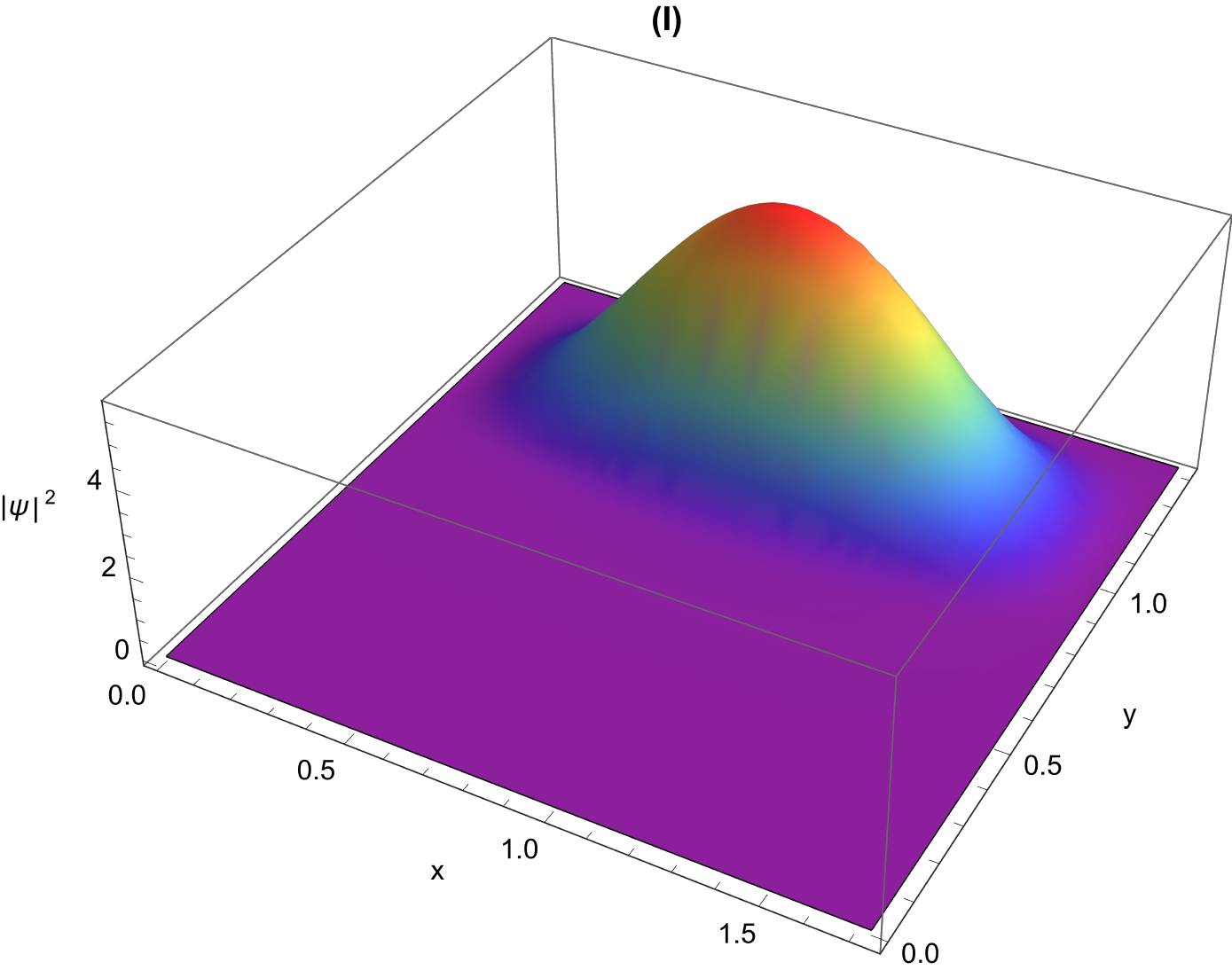}
 \includegraphics[width=0.3\columnwidth]{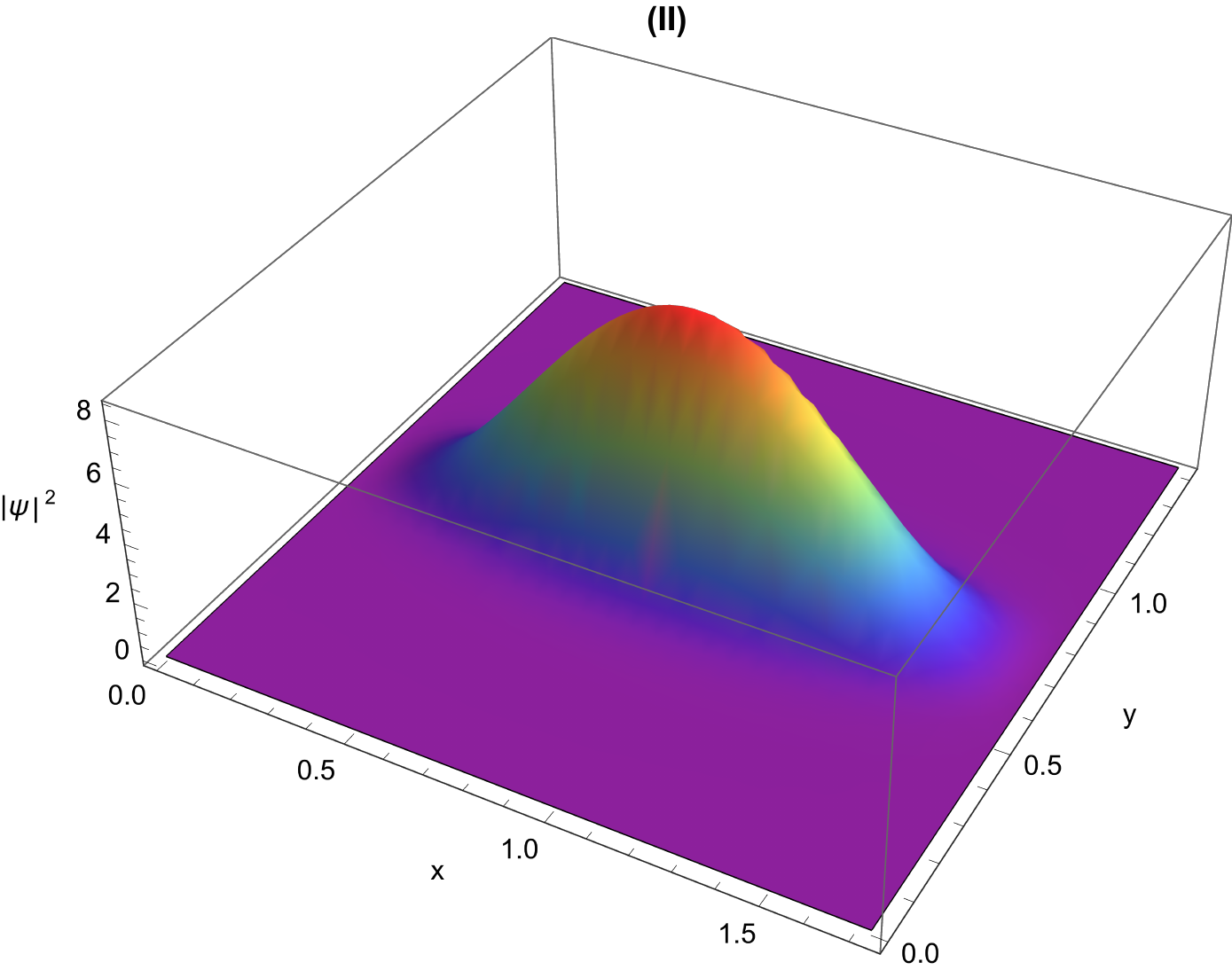}
 \includegraphics[width=0.3\columnwidth]{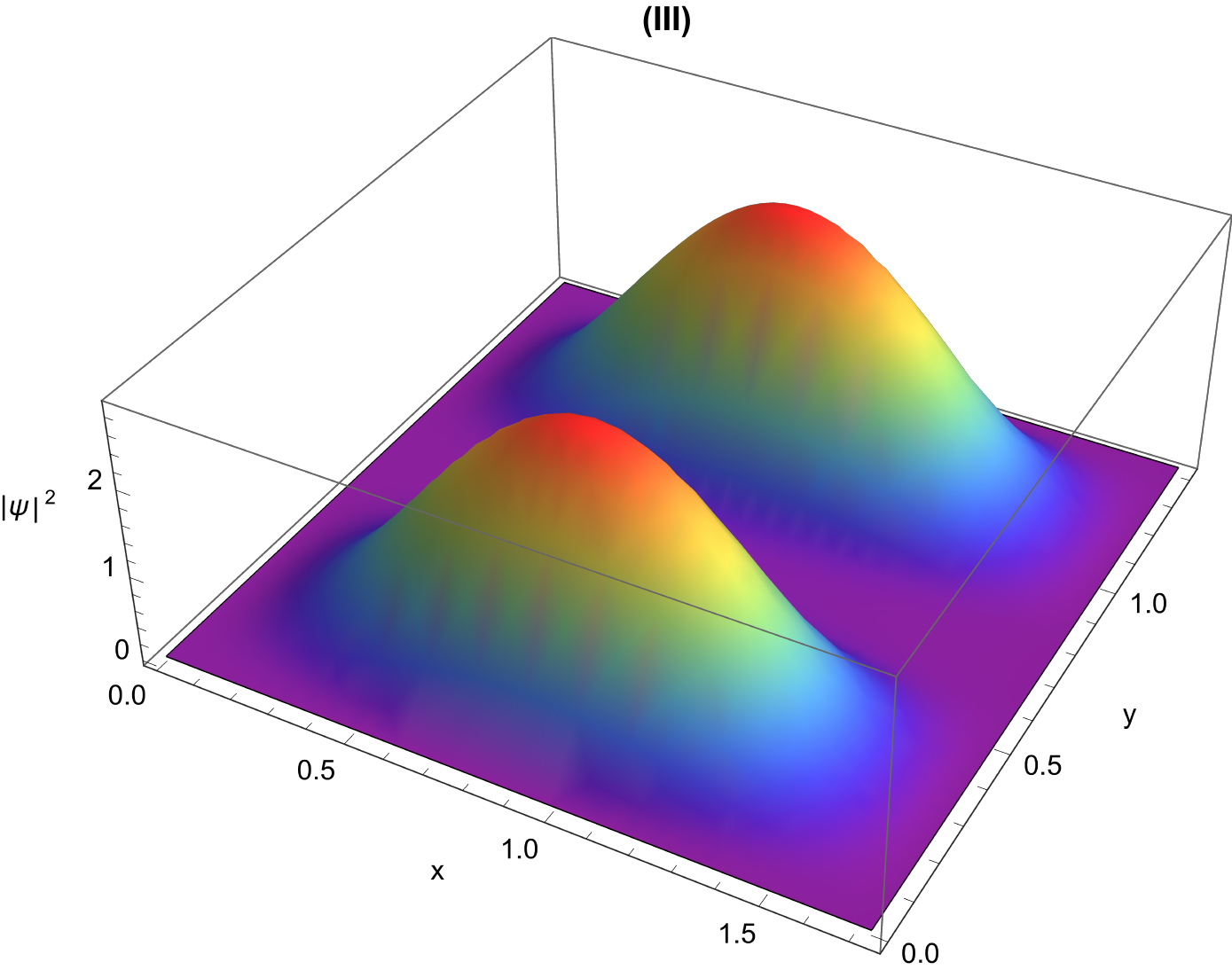}
 \includegraphics[width=0.3\columnwidth]{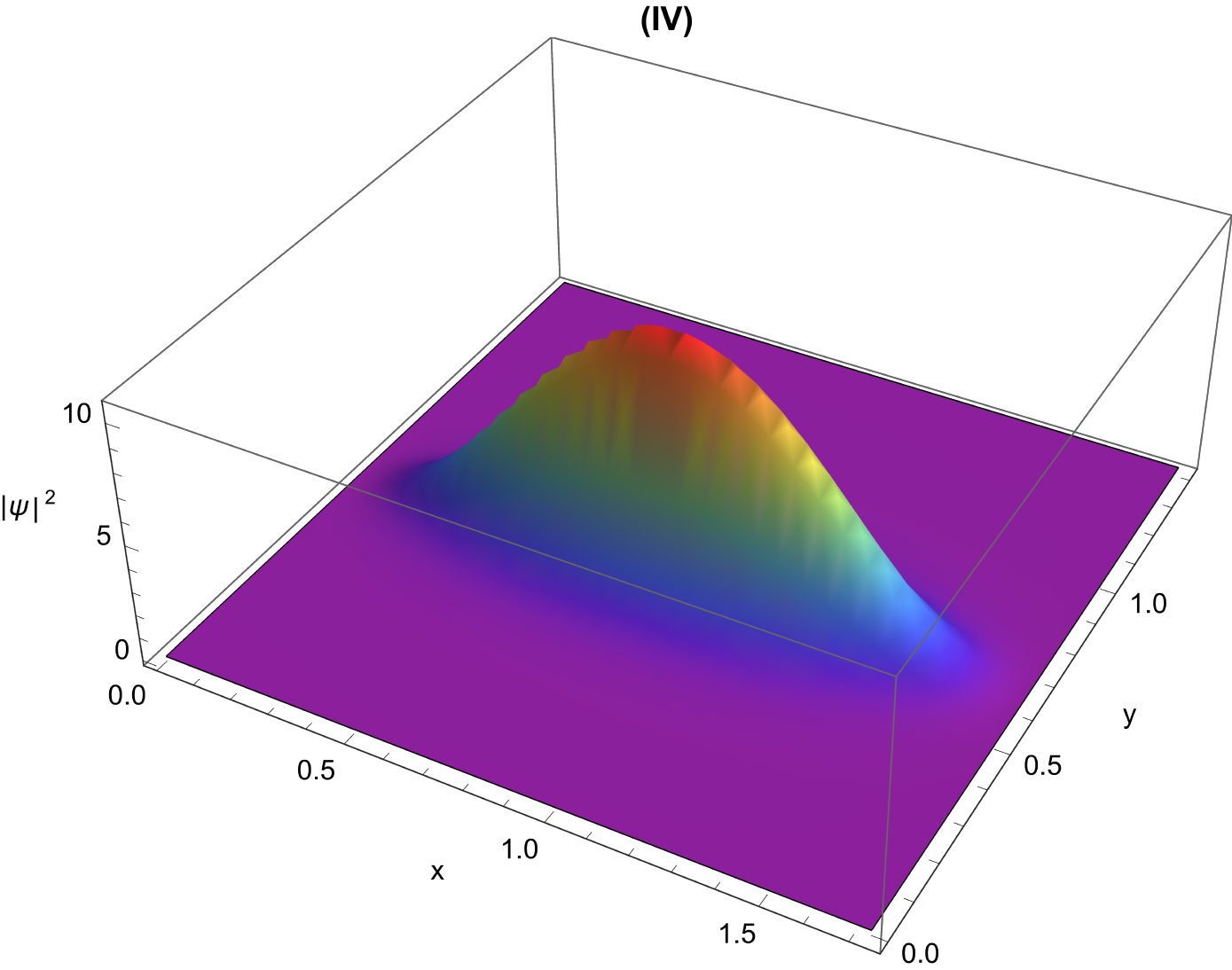}
 \includegraphics[width=0.3\columnwidth]{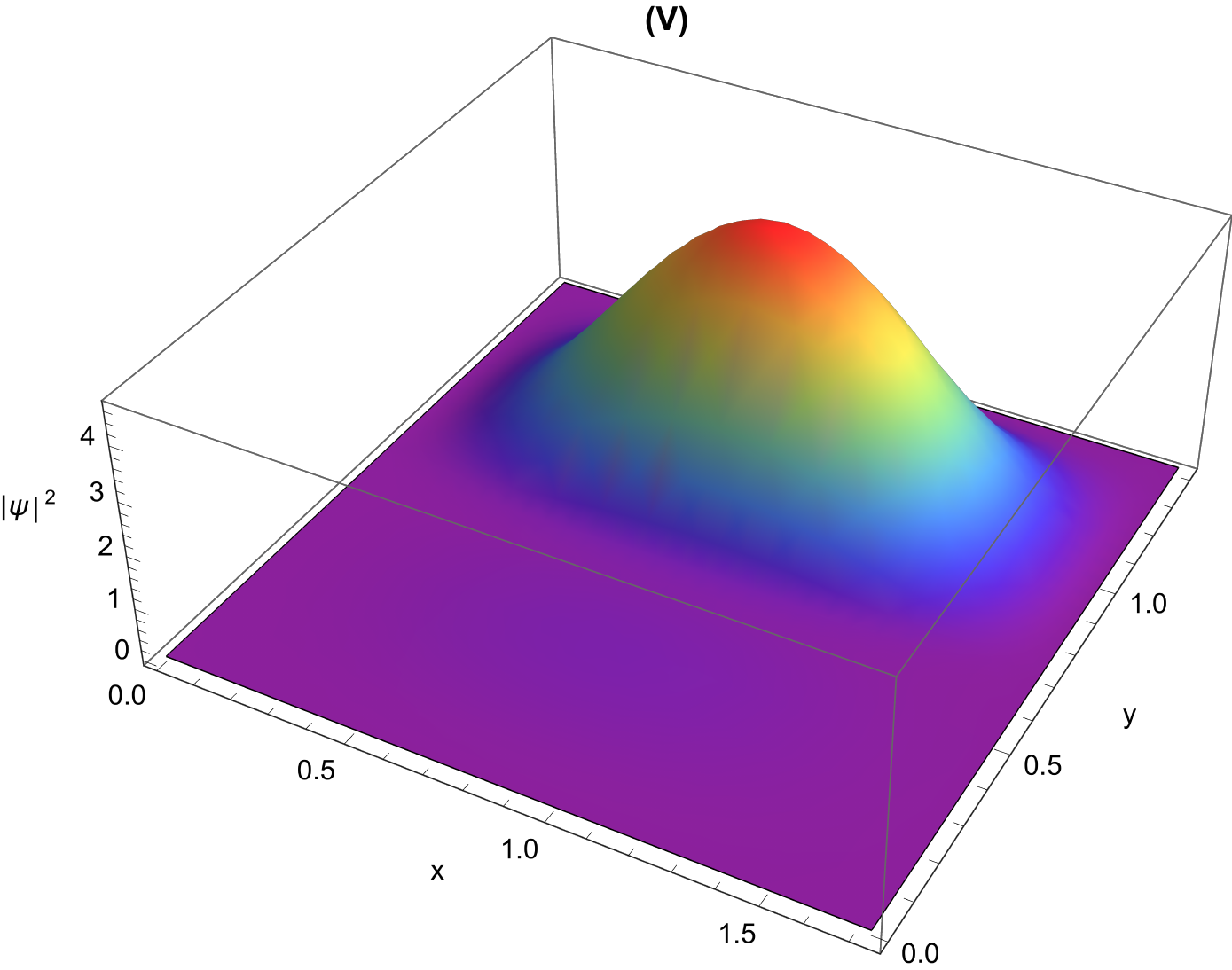}
 \includegraphics[width=0.3\columnwidth]{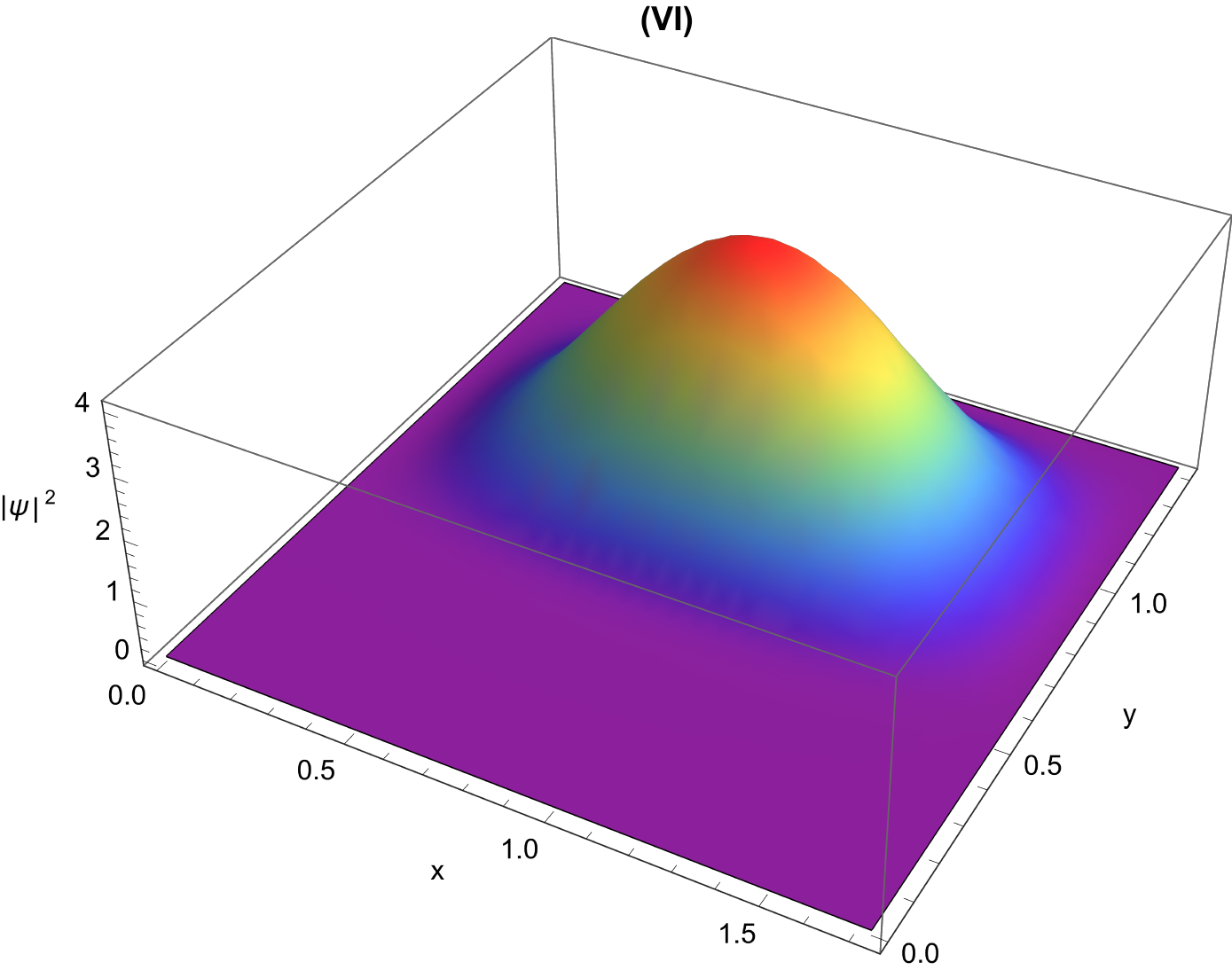}
 \includegraphics[width=0.3\columnwidth]{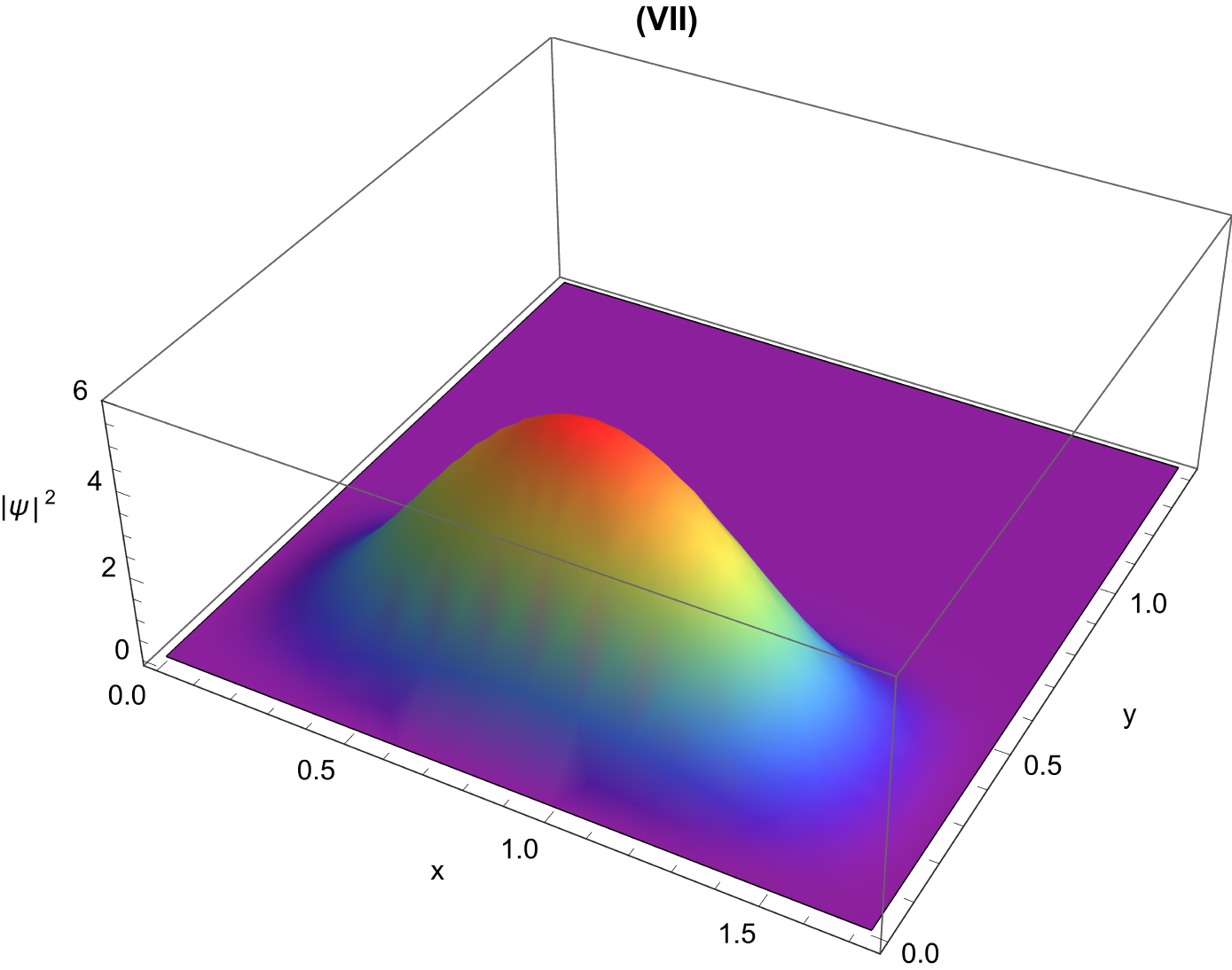}
  \caption{A collage of  normalized  ground states $2D$ probability densities $|\psi(x,y)|^{2}$  (plotted in arbitrary units), respectively corresponding to the seven
sets of striped, all real potentials tabulated in Table-\ref{Table-I} for the corresponding ground states (panels I through VII). The familiar ``Baseline" ($V=0$
everywhere inside the box) Figure precedes the profiles I-VII, for comparison.}
  \label{fig:Figure-2}
\end{figure}

Since, even the striped real valued $2$D potential distribution seems to have eluded attention, for an initial orientation, we present seven representative sets of all completely real-valued potentials with qualitatively different choices for $V_1, V_2, V_3, V_4$  (in Rydberg energy units).   The Hamiltonian operator on the left of Eq. (\ref{schro1}) is manifestly hermitian for this case, with no further symmetry requirements.  Table-\ref{Table-I} presents five lowest energy eigenvalues, of course all of them real, for each of the seven adumbrative cases (I-VII);  while Figure-\ref{fig:Figure-2} depicts the pertinent $2$D probability densities $|\psi(x,y)|^2$, where the  portrayals are individually normalized albeit drawn to arbitrary scales.   Figure-\ref{fig:Figure-2} succinctly conveys that the probability exhibits varying degrees of localization, region-wise.    Hereafter, distances in multiples of Bohr radii and energy values in Rydberg units will be understood.  The ``Baseline" refers to the standard problem where the inside potential identically zero, and as is evident, for which the $y$-separation also occurs; so that with $n^{(0)}_{x}=1$, $n_y$ also has a fixed value. 


\begin{table}[h]
\centering
\caption{Some typical, all-real-valued, seven sets of “striped” potentials (i.e. potential energies, in Ry) along the  four $y$-stripes (Figure-\ref{fig:Figure-1})  and the associated lowest five energy eigenvalues for a rectangular rigid box with $x$ length $a=\sqrt{3}$ and $y$-width  $b=\sqrt{2}$ in units of $a_B$, the Bohr radius.  The parameter $n^{(0)}_{x}=1$ so that the $y$-solution governs the number of peaks ({\it cf}. Eq.(\ref{psi})). For comparison, the  “Baseline” refers to the standard rigid box with the inside potential identically zero. }
\label{Table-I}
\begin{tabular}{|l|l|l|l|l|l|l|l|l|l|}
\hline
\multirow{2}{*}{Set} & \multicolumn{4}{|l|}{Potentials in $y$-strips (Ry)}       & \multicolumn{5}{|l|}{Lowest five energy eigenvalues (Ry)} \\ \cline{2-10}
                     & $V_1$ & $V_2$ & $V_3$ &  \multicolumn{1}{|l|}{$V_4$} &$E_1$& $E_2$& $E_3$ & $E_4$&$E_5$ \\ \cline{1-10}
\multirow{1}{*}{Baseline} & 0     & 0     & 0     & 0                           & 8.224        & 23.03       &47.70 & 82.25       & 126.7           \\ \cline{1-10}
I                                          & +100  & -100  & +100  & -100                        & -72.61       & -6.131       & 18.88       & 112.7       & 134.9      \\ \cline{1-10}
II                   & +100  & -100  & -100  & +100                        & -43.67       & 83.23        & 130.1       & 145.2       & 203.8      \\ \cline{1-10}
III                  & -100  & +100  & +100  & -100                        & -72.69       & -72.22       & -3.023      & 0.230      & 98.42      \\ \cline{1-10}
IV                   & +100  & -100  & +100  & +100                        & 12.56        & 119.9        & 132.9       & 171.9       & 211.8      \\ \cline{1-10}
V                    & -100  & +100  & -100  & -100                        & -80.60       & -72.45       & -32.74      & -1.211      & 46.75      \\ \cline{1-10}
VI                   & +100  & -100  & -100  & -100                        & -85.05       & -50.30       & 6.566       & 81.70       & 132.2      \\ \cline{1-10}
VII                  & -100  & +100  & +100 & +100                        & -72.45       & -1.436       & 101.7       & 121.6       & 161.1      \\ \cline{1-10}
\end{tabular}
\end{table}

The baseline energy levels are of course $\displaystyle E = E^{(0)}_{n_x,n_y} = \frac{\pi^2 \hbar^2}{2\mu}\Big(\frac{n^{2}_{x}}{a^2} + \frac{n^{2}_{y}}{b^2}\Big)$;
presented in Table-\ref{Table-I}: $n_x=n^{(0)}_{x} =1$ and $n_y$ values sweep through $1,2,3,4$ and $5$ where the baseline ground state ($n^{(0)}_{x}=1, n_y=1$)
probability density displays a single, flat peak (Figure-\ref{fig:Figure-2}). Although we shall not present in detail here, the expansion coefficients turn out contributing 
significantly at best only for the $n_y$ values ranging from $n_y\sim 1-8$. 
\subsection{Indroducing $\mathcal{PT}$ symmetric potentials}
Next, we appraise the effect of introducing $\mathcal{PT}$ symmetry in the striped regions.  We shall consider only the cases with nontrivial scenarios, where the participant states in $\mathcal{PT}$ sustenance and breakdown often will, but need not always imperatively involve, the ground-state.  Unless otherwise specified, we employ the eigenvalue-labeling convention from MATHEMATICA \cite{WOLF2003} where magnitude-wise the largest eigenvalues are labeled in diminishing order of their magnitude.  Labeling the energy-sequence ``{\it p}" in this reverse order, $\mathcal{PT}$  scenarios surrounding  $p =50, 49, 48, \cdots$ will be considered. 

\begin{figure}[h]
\centering

\includegraphics[width=0.5\columnwidth]{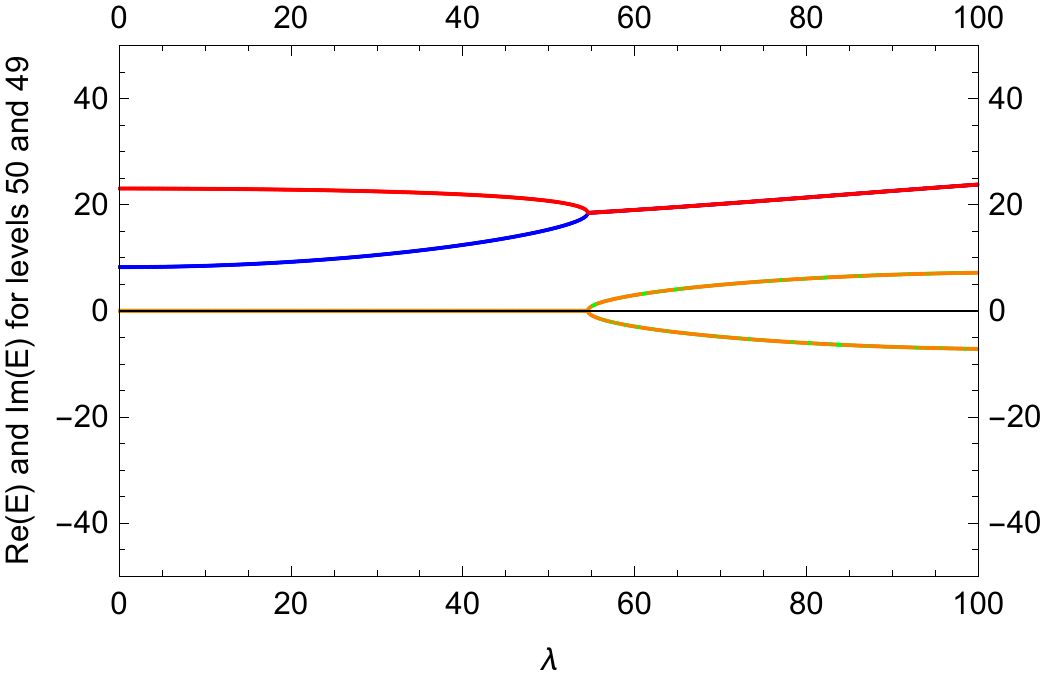}
  \caption{Plots for  $Re(E_p)$ and  $Im(E_p)$ parts of the eigenvalues $E_p$,  versus the  parameter $\lambda$ in $V_2=i\lambda=V^{*}_{3}$; $V_1=0=V_4$ 
for the levels  $p=50$ and $p=49$. Blue plot: $Re(E_{50})$, red plot: $Re(E_{49})$.  $\mathcal{PT}$ symmetry is retained, yielding real eigenvalues, 
up to the critical value $\lambda_c=54.5$ (``exceptional point") wherein the orange and green (overlapped) plots for $Im(E_{50})$ and  $Im(E_{49})$ respectively, are zero up to $\lambda_c$ beyond which the symmetry breaks down. Thereafter, the real parts merge together, 
while the imaginary parts are rendered equal and opposite, signifying occurrence of the eigenvalues in complex conjugate pairs. 
Energies in Ry, distances in Bohr radii. The quantum number $n^{(0)}_{x}$ has been clamped to unity.}
  \label{fig:ReImEigen1}
\end{figure}

\begin{figure}[h]
\centering
\includegraphics[width=0.5\columnwidth]{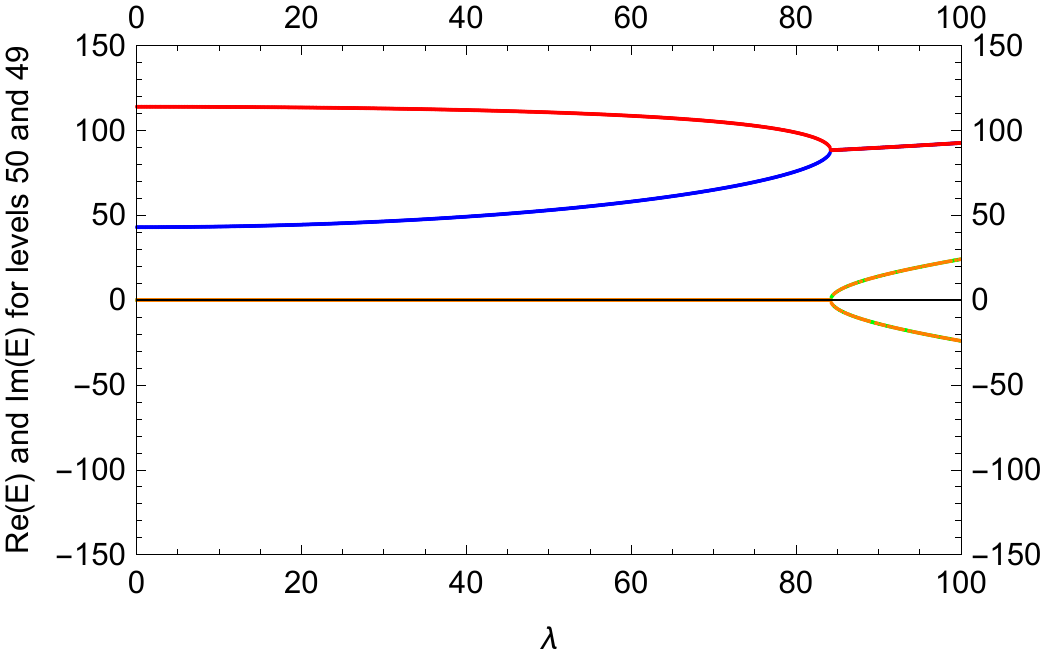}
  \caption{$V_2 = i\lambda=V^{*}_{3}, \lambda \rightarrow (0-100)$ and $n^{(0)}_{x}=1$. The bordering sectors however are now held at a higher potential $V_1=100=V_4$.  $Re(E_p)$ and $Im(E_p)$ are plotted versus $\lambda$, blue plot: $Re(E_{50})$, red plot: $Re(E_{49})$.  $\mathcal{PT}$ breakdown occurs at a higher 
$\lambda_c = 84.0$.  Likewise as in Figure-\ref{fig:ReImEigen1}, the orange and green (overlapped) plots represent $Im(E_{50})$ and $Im(E_{49})$.  
Again, beyond $\lambda_c$, the real parts merge and eigenvalues emerge only as complex conjugate pairs. }
  \label{fig:ReImEigen2}
\end{figure}

Consider the case where the two outer stripes are held at zero potential ($V_1=V_4=0$) while the inner two sectors are rendered $\mathcal{PT}$ symmetric, 
choosing ($V_2=i\lambda=V_{3}^{*}$) with the real, positive parameter $\lambda$ being continuously varied over the range $0$ through $100$.
 Figure-\ref{fig:ReImEigen1} plots the profiles for $Re(E_p)$ and $Im(E_p)$,  respectively the real  and imaginary-parts of the energy levels, for $p = 50$ 
as well as for $p = 49$ whose starting points happen to be truly the ground  and the first excited states.
  Initially, $\mathcal{PT}$ symmetry is retained as the two levels continue to remain conspicuously real valued and distinct, 
slowly start drifting toward each other and merge together at a characteristic critical threshold value $\lambda_c=54.5$, 
the exceptional `critical' point, precisely at which the breakdown is triggered.   For $\lambda>\lambda_c$, there is a complete $\mathcal{PT}$ breakdown 
for the two energy levels $E_{50}$ and $E_{49}$, which thereafter start occurring in complex conjugate pairs, in consonance 
with Eq.(\ref{PTeq2}), as demarcated in Figure-\ref{fig:ReImEigen1}. In similar fashion, raising the potentials in the two $y$-borderline 
sectors by setting $V_1=V_2=100$, and once again setting 
$V_2=i\lambda=V_{3}^{*}$ sweeping the range ($0-100$), the starting points here too  happen to be exactly the ground and the first excited states, 
the exceptional point shifts upward to $\lambda_c = 84.0$, exhibiting higher degree of 
retention of $\mathcal{PT}$ symmetry, as is evident from the transition depicted in Figure-\ref{fig:ReImEigen2}.

\begin{figure}[h]
\centering
\includegraphics[width=0.5\columnwidth]{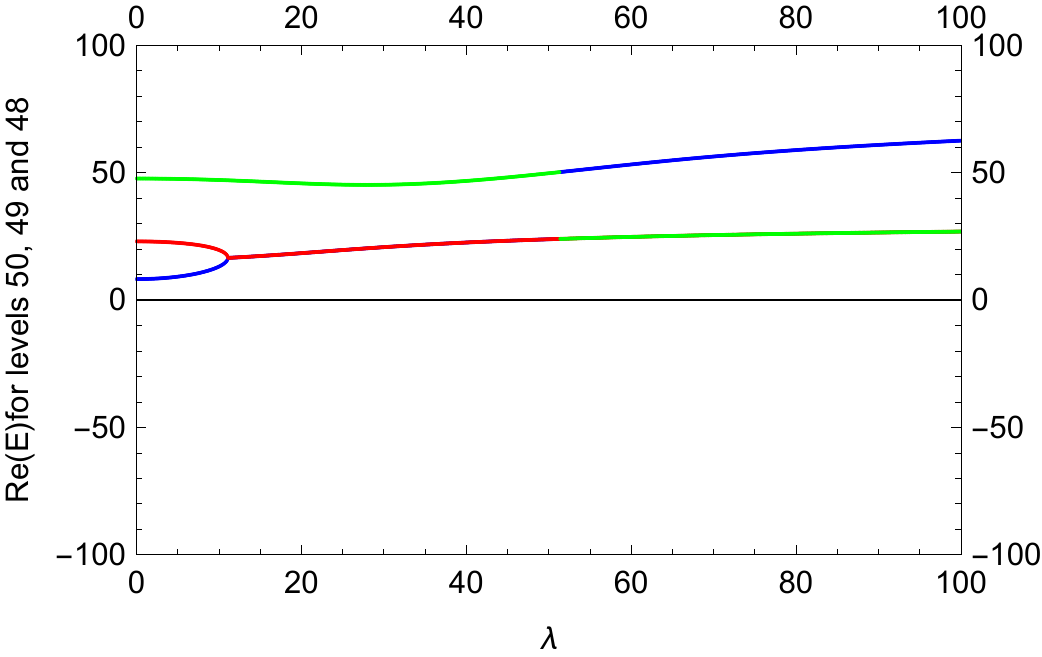}
  \caption{$Re(E_p)$ vs. $\lambda$ behavior for $p=50$ (blue), $p=49 $ (red) and $p=48$ (green), with the outer sectors imparted $\mathcal{PT}$ symmetric: $V_1=i\lambda
=V^{*}_{4}$; inside, $V_2=0=V_3$.  The range swept is: $\lambda \rightarrow (0-100)$. The quantum number $n^{(0)}_{x}=1$, as before.   A remarkable interplay of sustenance and breaking of $\mathcal{PT}$  occurs:  First breakdown for $p = 50, 49$ at $\lambda_{c_1}= 11.0$; $\mathcal{PT}$ restoration for $p = 50$ and a ‘crossover’ leading to breaking for $p=49$ and $48$,  at $\lambda = 51.0$  is evident.  See the text for fuller details;  {\it cf}. also the next figure (Figure-\ref{fig:ImEigen3}).}
  \label{fig:ReEigen3}
\end{figure}

\begin{figure}[h]
\centering
\includegraphics[width=0.5\columnwidth]{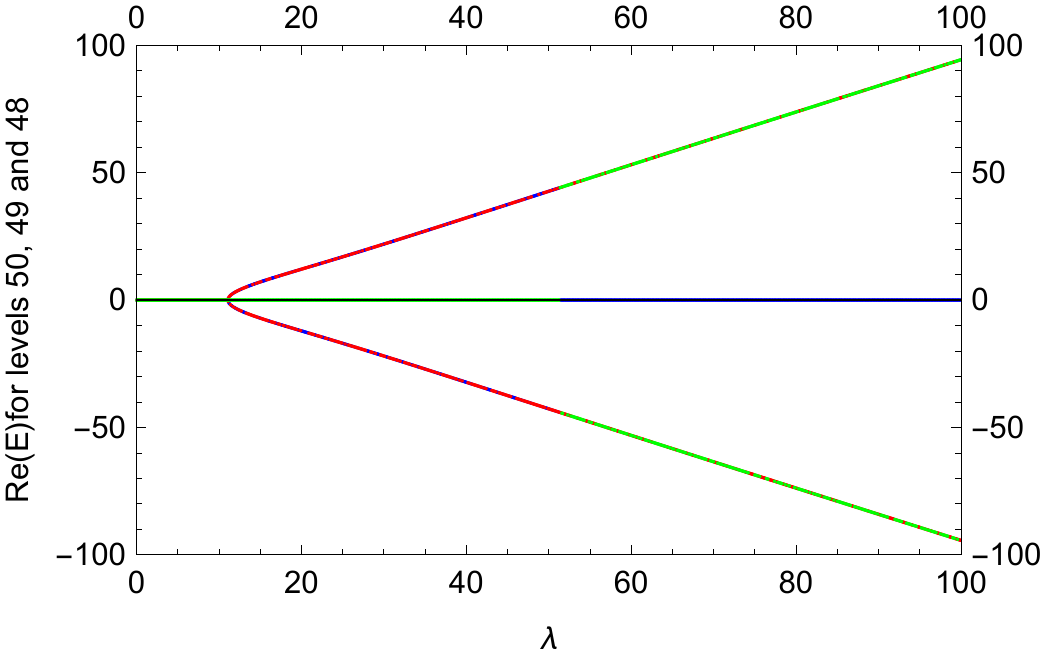}
  \caption{$Im(E_{p})$ vs. $\lambda$ behavior for $p=50$ (blue), $p=49$ (red) and $p=48$ (green).  The other control parameters are the same as 
those in Figure-\ref{fig:ReEigen3}.  On the extreme left, between the first ($\lambda_{c_1}=11.0$)  and the second ($\lambda_{c_2}=51.0$) criticality, the blue and red plots perfectly overlap, giving a magenta type hue.  Post $\lambda_{c_2}$, the eigenvalue $E_{50}$ becomes completely real; 
concomitantly preceding two eigenvalues emerge with $E_{49}=E^{*}_{48}$,  suffering a $\mathcal{PT}$  breakdown.}
  \label{fig:ImEigen3}
\end{figure}

Next, we scrutinize the response to $\mathcal{PT}$ symmetry by interchanging preceding two blocks: with now the outer sectors held non-hermitian $\mathcal{PT}$ symmetric: $V_1=i\lambda
=V^{*}_{4}$ and $\lambda\rightarrow(0-100)$ while, the interior set to zero: $V_2=0=V_3$   (the quantum number $n^{(0)}_{x}=1$, as before).  
A remarkable $\mathcal{PT}$ behavior then becomes manifest: First at the exceptional point $\lambda_{c_{1}}=1$ (first criticality),  a $\mathcal{PT}$ breakdown occurs for 
the two lowest two levels $p = 50$ (ground-state for $\lambda=0$), $p = 49$ (immediately succeeding excited state for $\lambda=0$).  
Thereafter, instead of continuing with the symmetry breach throughout the remainder of the $\lambda$-range, a {\it second} criticality emanates 
for the level $E_{50}$ at $\lambda_{c_2}=51.0$, where, surprisingly, not only there is a complete {\it restoration} of $\mathcal{PT}$ symmetry (i.e. being bestowed with completely real eigenvalues), for the (erstwhile-) lowest level $E_{50}$, but concomitantly, at precisely the same juncture, the next two higher states with $p = 49$ and $48$ 
“collude” together triggering a $\mathcal{PT}$ symmetry-breaking transition.  Further, this ‘crossover’ is particularly a smooth one, i.e. with no sudden kinks, as is vividly purported by the collages of the plots in Figures-\ref{fig:ReEigen3} and \ref{fig:ImEigen3}.  Note that at the point of second criticality $\lambda_{c_2}$, $E_{50}$ abruptly jumps upward and smoothly continues the sojourn of the state $E_{48}$, the latter steadily linking itself with the $\mathcal{PT}$  breakdown curve but now by coupling with the state $E_{49}$, with $Re(E_{48}), \ Re(E_{49})$ congruently merging together.  It is thus clear that, for real valued eigenvalues, unlike in adiabatic change of the parameters, the energy hierarchy is not necessarily preserved under $\mathcal{PT}$ transitions.  These counterintuitive tripartite peregrinations of the transiting participant states cannot be predicted {\it a priori}.

\begin{figure}[h]
\centering
\includegraphics[width=0.5\columnwidth]{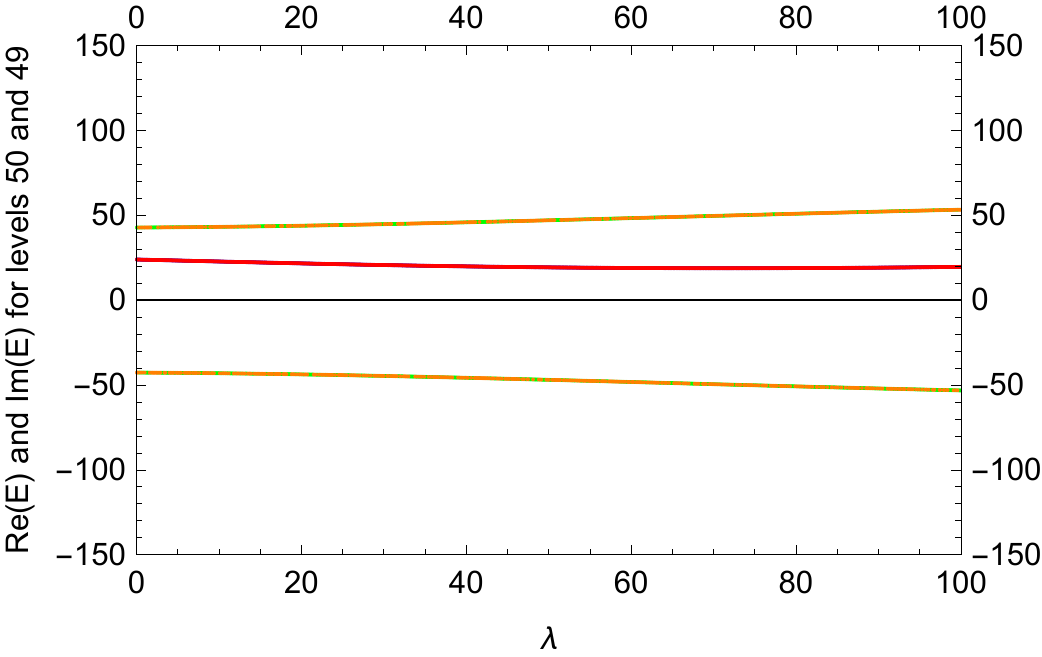}
  \caption{$Re(E_{50})=Re(E_{49})$ (both completely merged, in red) and $Im(E_{50})=-Im(E_{49}$) (orange-colored plots),  for a strong,  
fixed choice of  $V_1=50 \ i = V^{*}_{4}$  and varying $V_2 = i\lambda =V^{*}_{3}$  $\lambda\rightarrow (0-100)$.  $\mathcal{PT}$ breakdown, right from the start is evident throughout.}
  \label{fig:ReImEigen4}
\end{figure}

 Next, we empower the extreme two sectors with strong $\mathcal{PT}$ symmetry, vide $V_1=50 \ i = V^{*}_{4}$  and vary $V_2 = i\lambda =V^{*}_{3}$ with $\lambda\rightarrow(0-100)$ while, $n^{(0)}_{x}$  still kept locked  at unity, thus all the four regions are made non-hermitian $\mathcal{PT}$ symmetric.   In response, $\mathcal{PT}$ that is broken right in the beginning dismally fails to recover and continues to be breached, as the plots in Figure-\ref{fig:ReImEigen4} elucidate.

Incidentally, the present work turns out, in a sense, ‘a spatial counterpart’ to the interesting recent work by Hayrapetyan, Klevansky and G\"{o}tte \cite{HAYR2017}, who introduced in a spatially homogeneous optical medium, a time-dependent $\mathcal{PT}$ symmetric dielectric permittivity and demonstrated within the framework of classical electrodynamics, the consequent light amplification as well as attenuation.
\subsection{Probability density mutations}
\begin{figure}[h]
\centering
\includegraphics[width=0.4\columnwidth]{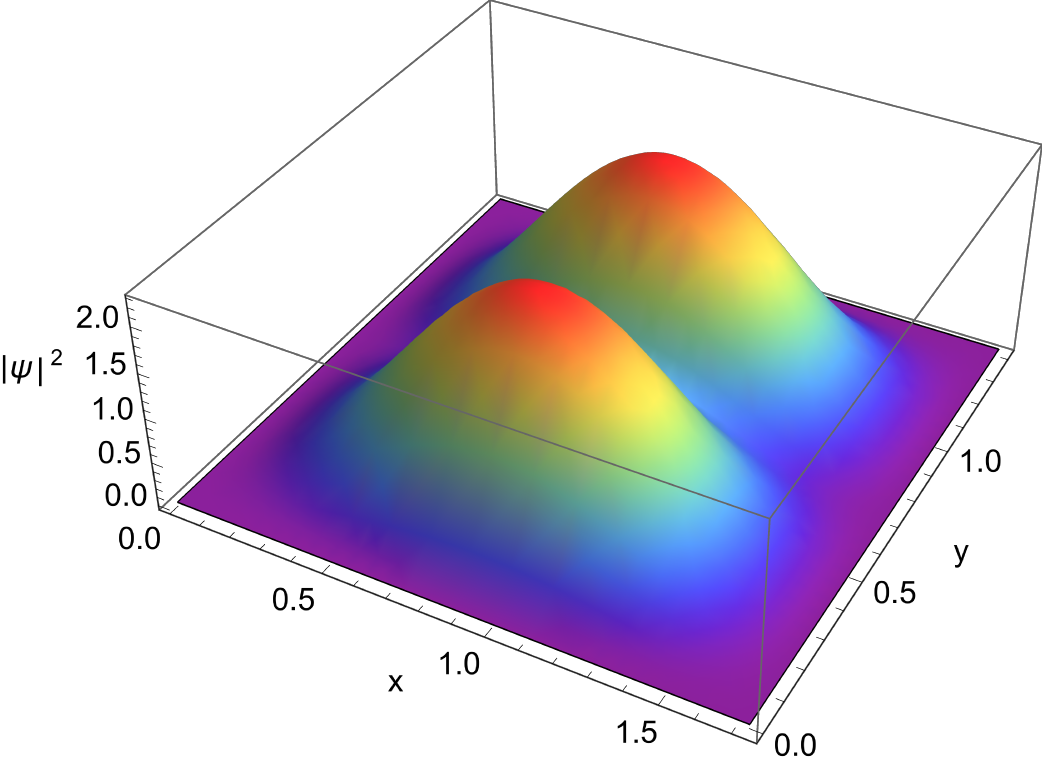}
\includegraphics[width=0.4\columnwidth]{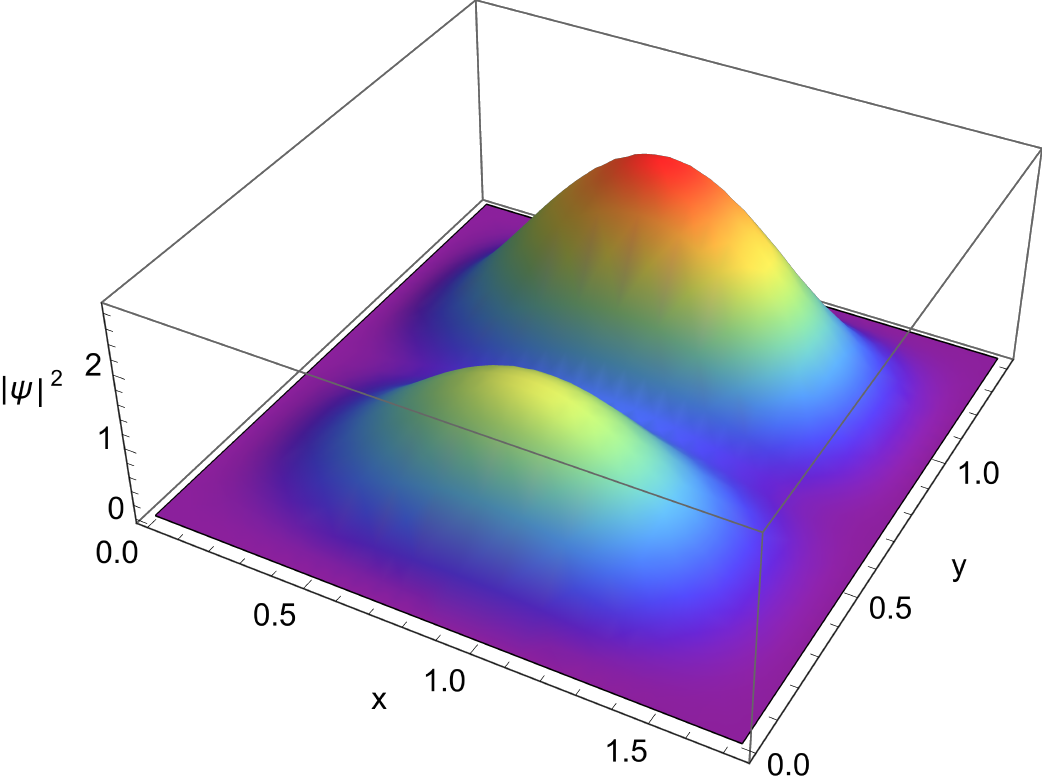}
  \caption{Pre (left) and Post (right) $\mathcal{PT}$ breaking Probability density  $|\psi(x,y)|^2$  for the ground-level $E_{50}$, with the same parameterization as in Figure-\ref{fig:ReImEigen1}; with $\lambda =54< 54.5=\lambda_c$ for pre-$\mathcal{PT}$ while, $\lambda =56> 54.5=\lambda_c$ for post-$\mathcal{PT}$ scenario. }
  \label{fig:psisqPT1}
\end{figure}

 Probability redistributions surrounding some typical $\mathcal{PT}$ transitions, are presented next.   Absolute squared of the ground-level wave function $|\psi(x,y)|^2$, pre (left plot of Figure-\ref{fig:psisqPT1}) and post(right plot of Figure-\ref{fig:psisqPT1}) $\mathcal{PT}$  breaking, presented to scale, reflects non-unitarity (in fact antiunitarity) of the composite $\mathcal{PT}$  operation evidenced from attenuation  in one of the peaks for the state $E_{50}$, after criticality ($\lambda_{c}=55.4$), for the same parameters as in Figure-\ref{fig:ReImEigen1}.

\begin{figure}[h]
\centering
\includegraphics[width=0.4\columnwidth]{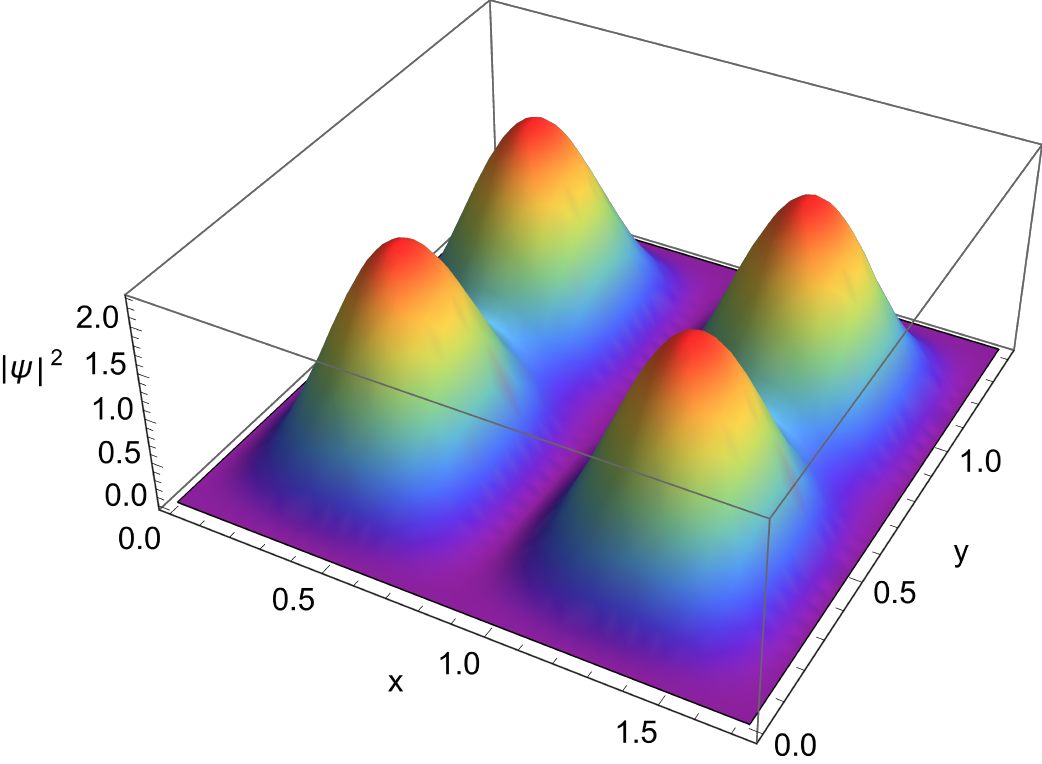}
\includegraphics[width=0.4\columnwidth]{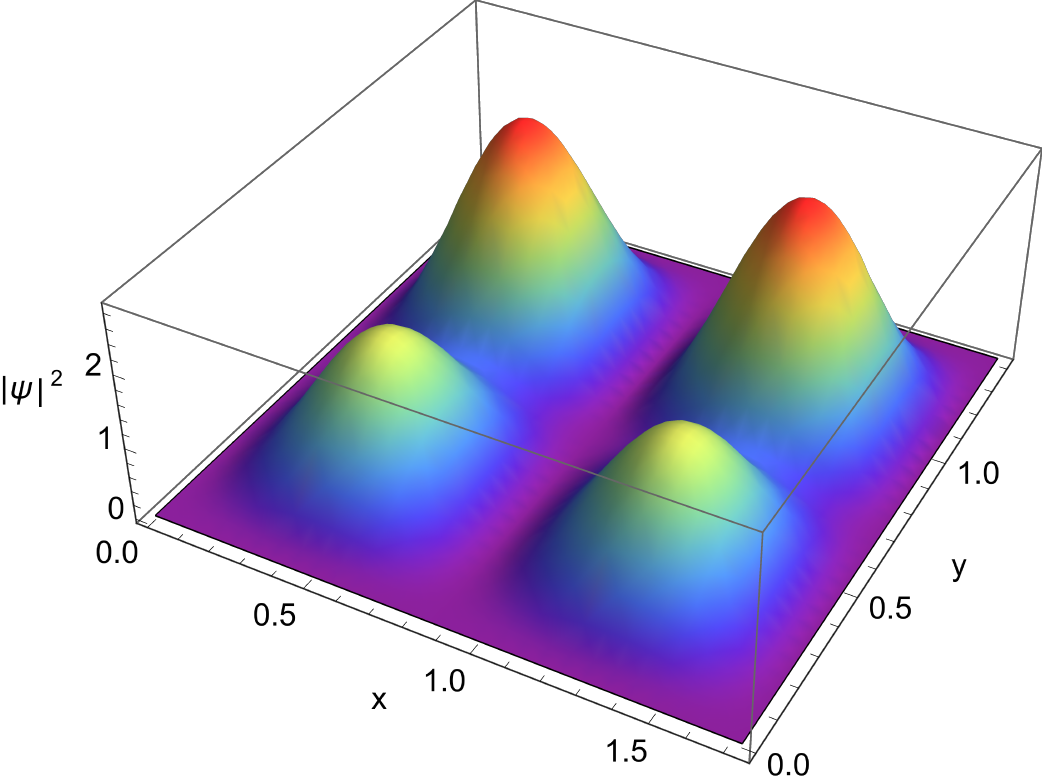}
  \caption{Pre (plot to the left) and Post (plot to the right)  $\mathcal{PT}$ breaking Probability density  $|\psi(x,y)|^2$  for the ground-level $E_{50}$,  $n^{(0)}_{x}=2$ with otherwise the same parameterization as in Figure-\ref{fig:ReImEigen1}; with $\lambda =54< 54.5=\lambda_c$  for pre-$\mathcal{PT}$ and $\lambda =56> 54.5=\lambda_c$  for post-$\mathcal{PT}$ scenario. }
  \label{fig:psisqPT3}
\end{figure}

Not surprisingly, this breakdown also prevails for $n^{(0)}_{x}=2$ where both the peaks deteriorate (Figure-\ref{fig:psisqPT3});  the factored out $x$-part of course having no direct role to play in the present $\mathcal{PT}$ gamut. 

Incidentally, with a small number of partitions as portrayed in Figure-\ref{fig:Figure-1}, it is straightforward to approach the present problem {\it directly}.  In the respective regions, one has the following forms  for the wave functions:
\bea
\psi(x,y) &\sim& A \sin\Big(\frac{n^{(0)}_{x}\pi x}{a}\Big)\sin(k_1 y)  \ ; \qquad\qquad \ \ \  \quad \ \ \text{Region-I}\nn\\
\psi(x,y) &\sim&  \sin\Big(\frac{n^{(0)}_{x}\pi x}{a}\Big)[B\sin(k_2 y)+C\cos(k_2 y)]  \ ; \ \text{Region-II}\nn\\
\psi(x,y) &\sim&  \sin\Big(\frac{n^{(0)}_{x}\pi x}{a}\Big)[F\sin(k_3 y)+G\cos(k_3 y)]  \ ; \ \text{Region-III}\nn\\
\psi(x,y) &\sim&  H \sin\Big(\frac{n^{(0)}_{x}\pi x}{a}\Big) \sin[k_{4}(y-b)]  \ ; \qquad\qquad \     \text{Region-IV} 
\eea 
the constant coefficients  determinable by matching only the  $y$-factors of wave functions and their derivatives at the respective interfaces, 
with the constants $\{k_i\}_{i=1}^{4}$ prescribed by the connection 
\beq
\frac{\hbar^2}{2\mu}k_{i}^{2} = E-V_{i} + \frac{\pi^2 \hbar^2}{2\mu}\Big(\frac{n^{(0)}_{x}}{a}\Big)^2; \ \forall i = 1,\cdots,4
\eeq
Note that the energy eigenvalues $E$ are unknown to begin with; the task is then tantamount to solving for the eigenvalues and the wave functions, the latter through determination of the coefficients.  However, this approach leads to a formidable set of coupled transcendental equations tractable only through numerical techniques; the situation getting increasingly aggravated as the number of partitions increases. Moreover, if one did venture to solve these simultaneously, one can in principle obtain but only a {\it specific} energy eigenvalue at a time.  For instance, even in the completely Hermitian case (all real potentials) the energy hierarchy will be indiscernible under this direct approach, as it is unclear {\it a priori}, exactly which eigenvalue - ground or excited - has resulted.  The present matrix approach, on the other hand offers a holistic method that unequivocally yields the complete eigenspectrum.    
\subsection{ $\mathcal{PT}$ symmetric electrostatic potential}
$\mathcal{PT}$  symmetry that hinges on an intricate balance between gain and loss mechanisms  is also investigated furnishing a fictitious electric field with a backdrop of the striped box, for the particle moving in the box  possessing  an electric charge $q$.  Consider a uniform electric field, say along the $y$-direction, $\vec{F}=F_{o}\hat{j}$
 furnished inside the box, and zero outside.  This engenders the electrostatic potential (energy), once again with a manifest explicit dependence only on the $y$-coordinate (apart from a trivial additive constant set to zero here):  $V(x,y)\equiv V(y)= -\alpha(y-b/2)$  with  $\alpha=qF_{o}$.  A real $\alpha$ imparts reality to $V$, which is manifestly antisymmetric around $y = b/2$, but still rendering the Hamiltonian hermitian; whilst a pure imaginary $\alpha=i\lambda, \ \lambda \in \mathbb{R}$ , generates a non-hermitian but a truly $\mathcal{PT}$ symmetric Hamiltonian associated with the corresponding electric field.  Following a similar chain of arguments as was done surrounding Eqs. (\ref{schro1}) through (\ref{mateq2}) above for the striped potential distribution,  the electric field gives rise to simple prescriptions  for the corresponding $M$- matrix, $M^{el}$, as described below. 

Since $V$ corresponding to the electric field is clearly antisymmetric in the $y$-coordinate around the median, 
the pertinent integrals all vanish for diagonal part of $M^{el}$, resulting strikingly, in exactly the same contribution as that of the rigid box with identically zero inside potential, 
\beq
M^{el}_{n'_{y},n_{y}} = \frac{\pi^2 \hbar^2}{2\mu}\Big[\frac{(n^{(0)}_{x})^2}{a^2}+\frac{(n_{y})^2}{b^2}\Big] \label{matE}
\eeq
for the diagonal elements ($n'_{y}=n_{y}$), while the off-diagonal portion can be written as
\beq
M^{el}_{n'_{y},n_{y}} = -\frac{\alpha b}{\pi^2}\Big[\frac{1}{(n'_{y}+n_y)^2} - \frac{1}{(n'_{y} - n_{y})^2}\Big] (1-(-1)^{n'_{y}+n_y}) \label{matE1}
\eeq
hence, contributing only for odd combinations of  $n'_{y}+n_y$; vanishing otherwise. 

\begin{figure}[h]
\centering
\includegraphics[width=0.4\columnwidth]{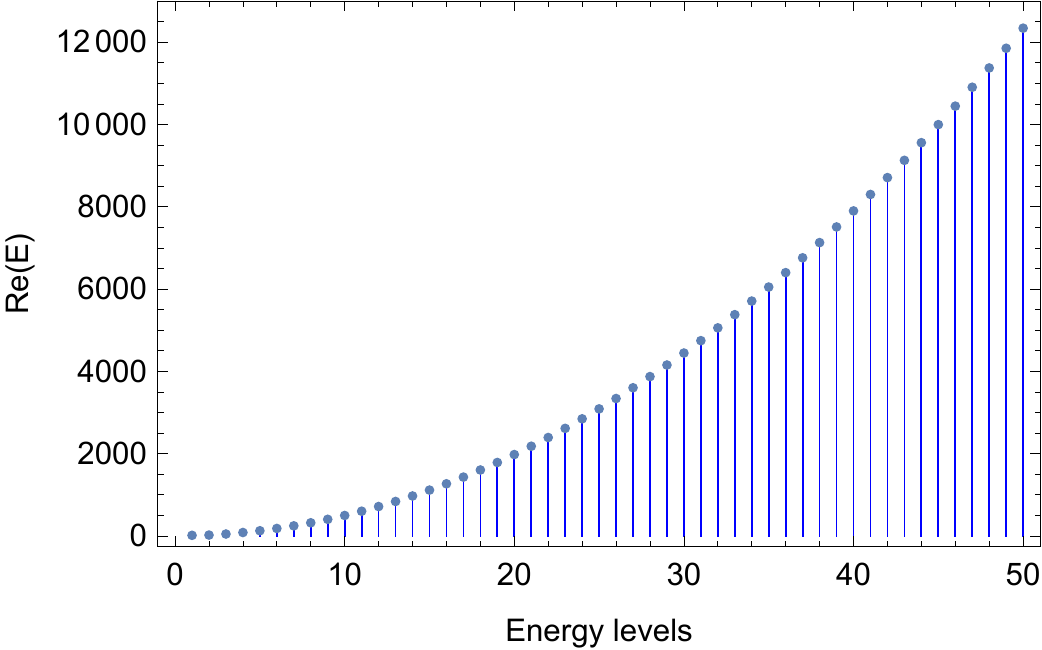}
\includegraphics[width=0.4\columnwidth]{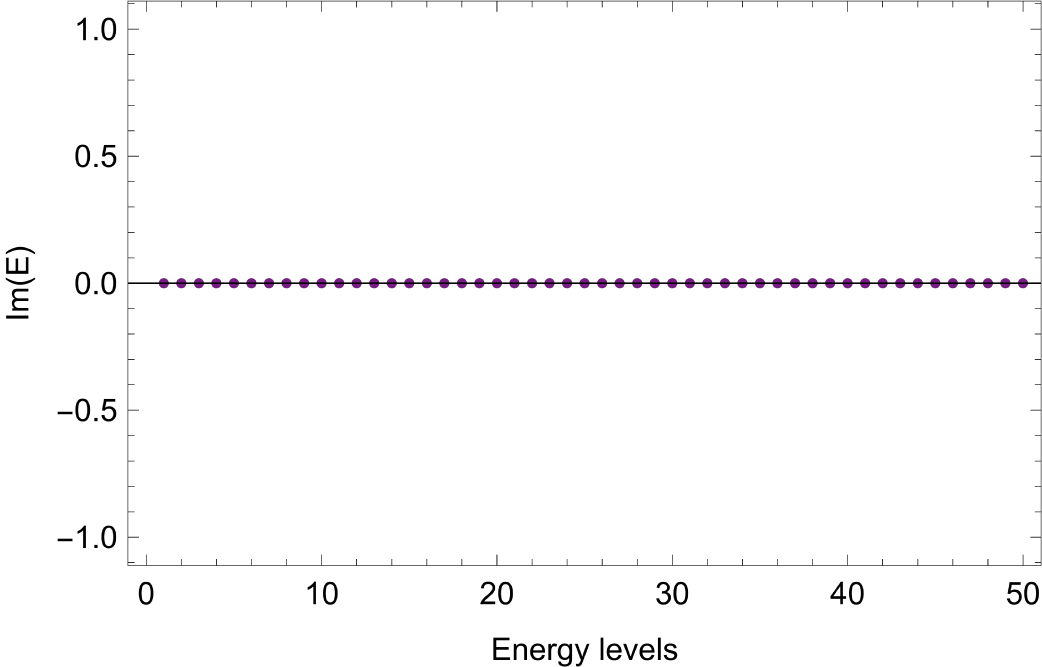}
\includegraphics[width=0.4\columnwidth]{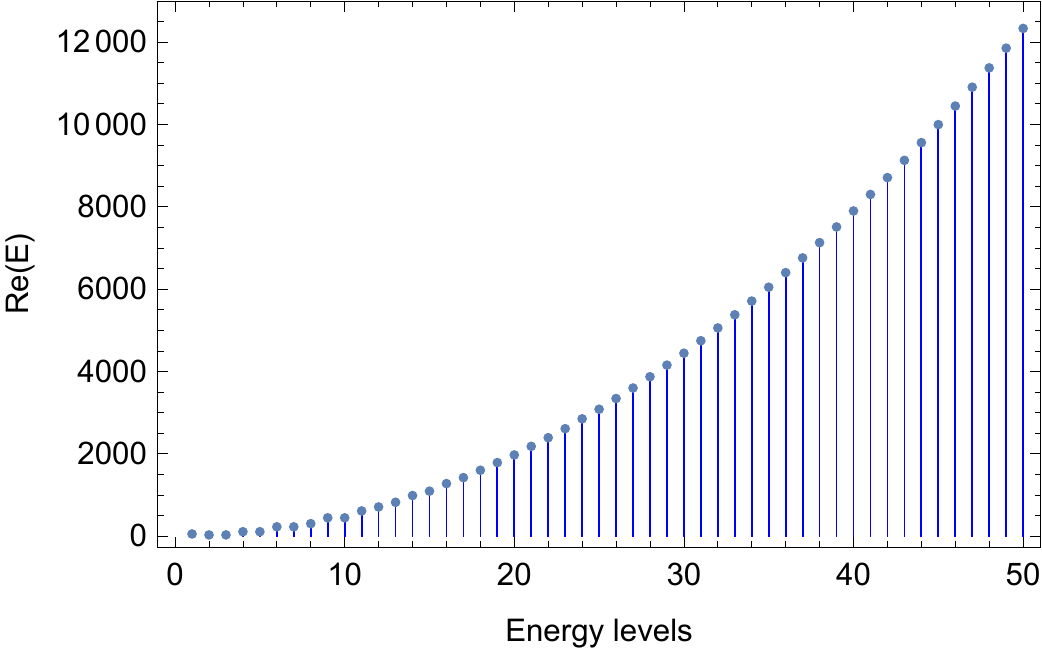}
\includegraphics[width=0.4\columnwidth]{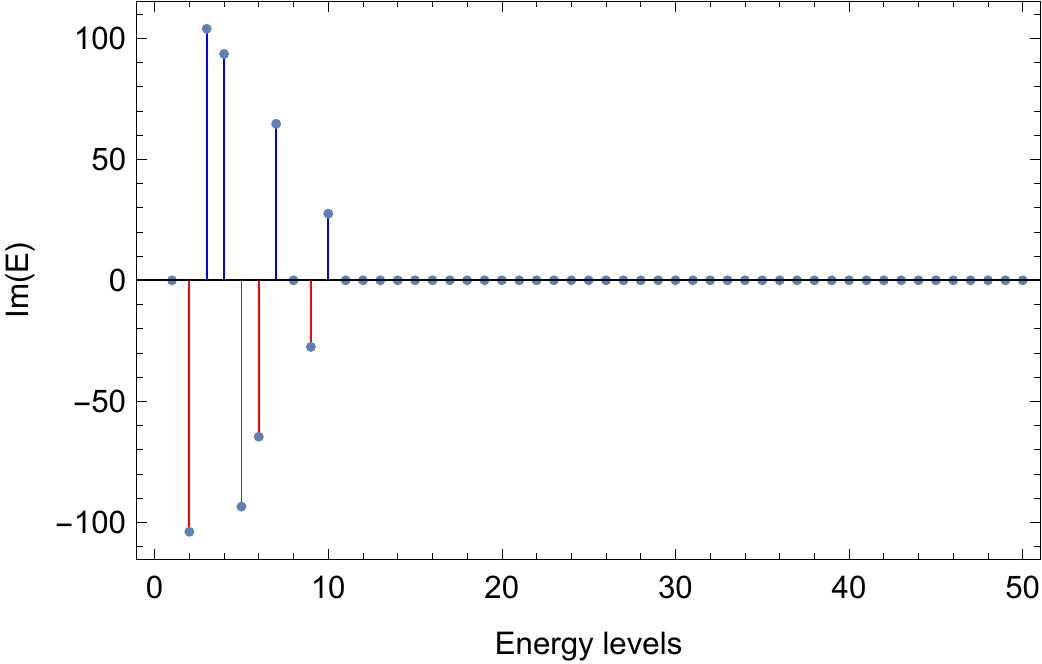}
  \caption{Top panels: The Real (Left) and Imaginary (Right) parts of the eigenenergies for a $\mathcal{PT}$ symmetric Electric Field $\alpha=20i$, with a $\mathcal{PT}$ symmetric background:$V_1=5i=V^{*}_{4}$ and $V_2=-5i=V^{*}_{3}$.  $\mathcal{PT}$  is preserved throughout: the eigenvalues are all real. Bottom panels:  $\alpha=20i$, with a stronger  $\mathcal{PT}$ backdrop, $V_1=100i=V^{*}_{4}$ and $V_2=-100i=V^{*}_{3}$, the symmetry breaks for a few states.  For convenience and to guide the eye, the eigenvalues have been labeled from 
$1$ through $50$ in increasing order of their magnitude.  }
  \label{fig:PTelectric}
\end{figure}

 If the electric field contribution is taken in conjunction with that of the striped potential, for the combined problem, only the off-diagonal part, Eq. (\ref{mateq2}), must simply be {\it augmented} by the expression on the right of Eq. (\ref{matE1}) above.   Of the number of qualitatively diverse combinations of the striped potential annexed by the uniform electric field, we shall herein present only a couple of salient aspects, summarized in Figure-\ref{fig:PTelectric}.  For a $\mathcal{PT}$  Symmetric Electric Field borne by $\alpha=20i$ along with a ‘moderate’ $\mathcal{PT}$  symmetric backdrop of $V_1=5i=V^{*}_{4}$ and $V_2=-5i=V^{*}_{3}$, $\mathcal{PT}$ is preserved throughout (top panels, Figure-\ref{fig:PTelectric}) as the resulting energy levels are all real.  Enhancing the background $\mathcal{PT}$ strength with the choice  $V_1=100i=V^{*}_{4}$ and $V_2=-100i=V^{*}_{3}$  the symmetry, albeit preserved for the ground-level, does break off for a handful, though not exhaustively all, of the higher ones (bottom panels, Figure-\ref{fig:PTelectric}), with emergence of equal and opposite imaginary parts. The $Re(E)$ plots are similar but not identical and practically indistinguishable on the scale plotted.  There thus are interesting trade-offs between $\mathcal{PT}$ symmetry transitions caused by the fictitious electric field and of the striped potential. 

\section{Conclusions}
 In summary, the present article derives exact quantum mechanical solutions for a $2$D rectangular rigid box with a piecewise constant set of potentials in parallel, contiguously connected rectangular sectors or ‘stripes’ that have their widths the same as that of the box, forming a special type of quantum-billiards.  These potentials are studied to gauge the response of a scalar particle both under hermitian and non-hermitian yet $\mathcal{PT}$ symmetric attributes ascribed, and combinations thereof.  In the latter, several interesting$\mathcal{PT}$ symmetry sustenance and breakdown transition scenarios emerge, including an intriguing reinstating of the broken symmetry for an energy level, concomitant with a smooth crossover of breakdown transition transferred to higher states.  $\mathcal{PT}$ symmetry that hinges on an intricate balance between gain and loss mechanisms is also investigated with a fictitious electric field furnished against a backdrop of the striped box.  It is gratifying all that these spectacular systems are amenable to an exact quantum mechanical treatment.  It would be instructive  to investigate the temporal evolution of 
an initially localized wave-packet (with compact support) within the box, as well as under exotic electromagnetic fields, within the $\mathcal{PT}$ symmetric framework.   These studies are currently under investigation.  
\vskip0.2cm
{\bf \large{Acknowledgments}}\\
The authors are immensely indebted to Professor Dr. P. Durganandini for several interesting discussions on the current theme, and her encouragement throughout.  SK is supported by University Grants Commission's Faculty Recharge Programme (UGC-FRP), Govt. of India,  New Delhi, India.

\end{document}